\documentclass[preprint,authoryear]{elsarticle}
\usepackage{natbib}
\usepackage{url} 
\usepackage{graphicx}
\graphicspath{ {figures/} }

\usepackage[usenames,dvipsnames]{color}
\usepackage{amsmath, amssymb}
\usepackage{amsfonts}
\usepackage{bm}
\usepackage[utf8]{inputenc}
\usepackage{times}
\usepackage{hyperref}
\hypersetup{
	colorlinks,
	citecolor=black,
	filecolor=black,
	linkcolor=black,
	urlcolor=black
}
\usepackage[utf8]{inputenc} 
\usepackage[T1]{fontenc} 

\usepackage[autostyle=true]{csquotes} 
\usepackage[a4paper,width=150mm,top=25mm,bottom=25mm]{geometry}
\usepackage{wrapfig}
\usepackage[final]{pdfpages}
\usepackage{hyperref}
\hypersetup{
	colorlinks,
	citecolor=black,
	filecolor=black,
	linkcolor=black,
	urlcolor=black
}

\usepackage{xparse}
\usepackage{physics}

\newcommand{\pd}[2]{\frac{\partial #1}{\partial #2}}

\usepackage[usenames,dvipsnames]{color}
\usepackage{newtxtext, newtxmath}

\begin{document}

\begin{frontmatter}
\title{A comprehensive study of nonlinear perturbations in the dynamics of planar crack fronts}

\author[1]{Itamar Kolvin\corref{cor1}}
\ead{ikolvin@gatech.edu}

\author[2]{Mokhtar Adda-Bedia}

\cortext[cor1]{Corresponding author}

\affiliation[1]{organization = {School of Physics, Georgia Institute of Technology},
addressline = {837 State St NW},
city = {Atlanta, GA},
postcode = {30332}, 
country = {USA} }
\affiliation[2]{organization = {Laboratoire de Physique, CNRS, ENS de Lyon, Universit\'e de Lyon},
city = {Lyon},
postcode = {F-69342}, 
country = {France} }
	
\begin{abstract}
The interaction of crack fronts with asperities is central to fracture criteria in heterogeneous materials and for predicting fracture surface formation. It is known how dynamic crack fronts respond to small, 1\textsuperscript{st}-order, perturbations. However, large and localized disturbances to crack motion induce dynamic and geometric nonlinear effects beyond the existing linear theories. Because the determination of the 3D elastic fields surrounding perturbed crack fronts is a necessary step toward any theoretical study of crack front dynamics, we develop a 2\textsuperscript{nd}-order perturbation theory for the asymptotic fields of planar crack fronts. Based on previous work, we consider two models of fracture: (1) Fracture in a scalar elastic solid which is an analog of antiplane shear fracture (Mode~III). In this model, the near-crack fields are obtained via matched asymptotic expansions. (2) Tensile Mode I fracture, in which a self-consistent expansion is used to resolve the fields near the crack front. These methods can be readily extended to higher perturbation orders. The main results of this work are the \textit{explicit} 2\textsuperscript{nd}-order expressions of the \textit{local} dynamic energy-release-rates for arbitrary perturbations of straight fronts. The formulae recover the known energy-release-rates of curved quasi-static fronts and of simple 2D cracks. We show that the expressions are separable as a product of a dynamical prefactor that only depends on the instantaneous local normal front velocity, and a history functional that integrates past front configurations. To gain insight, the energy-release-rates in the two models are computed for a traveling wave perturbation. While similar at low wave velocities, the two theories behave differently for fast waves. In scalar elasticity, the 2\textsuperscript{nd}-order contributions are always sub-dominant. However, in the Mode I theory, the 2\textsuperscript{nd}-order correction becomes the dominant term at the crack front wave velocity, where the 1\textsuperscript{st}-order term is zero. We discuss employing the energy-release-rate expressions to predict crack front dynamics via energy balance with dissipation.
\end{abstract}


\end{frontmatter}

\section{Introduction}

The prediction of crack propagation in heterogeneous media is a central problem in fracture mechanics~\citep{ponsonDepinningTransitionFailure2009,demeryMicrostructuralFeaturesEffective2014,steinhardtHowMaterialHeterogeneity2022,schmittbuhlDirectObservationSelfAffine1997,ponsonDepinningTransitionFailure2009,chopinMorphologyDynamicsCrack2015,lebihainEffectiveToughnessPeriodic2020,albertiniEffectiveToughnessHeterogeneous2021,rochDynamicCrackFrontDeformations2023,stanchitsHydraulicFracturingHeterogeneous2015,lubomirskyQuenchedDisorderInstability2023,cochardPropagationExtendedFractures2024}, in designing advanced materials~\citep{guptaToughDuctileArchitected,shaikeeaToughnessMechanicalMetamaterials2022,mirkhalafOvercomingBrittlenessGlass2014,xiaTougheningAsymmetryPeeling2012}, and in frictional and earthquake mechanics~\citep{latourEffectFaultHeterogeneity2013,lebihainEarthquakeNucleationFaults2021,gounonRuptureNucleationPeriodically2022,svetlizkyClassicalShearCracks2014,rayEarthquakeNucleationFaults2017,bayartRuptureDynamicsHeterogeneous2018,bedfordFaultRockHeterogeneity2022}. Computational frameworks for 3D dynamic fracture, such as the spectral boundary integral method~\citep{geubelleSpectralMethodThreedimensional1995,rochCRackletSpectralBoundary2022,rochDynamicCrackFrontDeformations2023}, phase-field simulations \citep{ponsHelicalCrackfrontInstability2010,henryFractographicAspectsCrack2013a,Chen.15,bleyerMicrobranchingInstabilityPhasefield2017,henryLimitationsModellingCrack2019,goswamiPhysicsinformedVariationalDeepONet2022}, and atomistic models \citep{heizlerMicrobranchingModeIFracture2015,mollerInfluenceCrackFront2015,buehlerModelingAtomisticDynamic2022}, perform the intensive task of numerically determining the elastodynamic fields. A complementary approach, pursued in this work, reduces the complexity of the 3D problem to the motion of 1D crack fronts. Cracks in brittle materials are described by the theory of Linear Elastic Fracture Mechanics (LEFM). The cornerstone of LEFM is that the near-crack region is dominated by a universal $r^{-1/2}$ stress field characterized by a single intensity factor. The stress intensity factor (SIF), then, determines the crack dynamics. This approach was adopted in several works \citep{riceThreedimensionalPerturbationSolution1994,willisDynamicWeightFunctions1995,willisThreedimensionalDynamicPerturbation1997,Ramanathan.97,morrisseyCrackFrontWaves1998,Morrissey.00,norrisMultiplescalesApproachCrackfront2007,willisCrackFrontPerturbations2013,adda-bediaDynamicStabilityCrack2013} which considered small perturbations to straight crack fronts. The usefulness of this approach was demonstrated by the discovery of crack front waves~\citep{Ramanathan.97,morrisseyCrackFrontWaves1998,Morrissey.00}, which were later observed experimentally \citep{sharonPropagatingSolitaryWaves2001,sharon2002crack,Fineberg.03,Sagy.04}. When applied to mixed mode loading configurations, this approach yielded possible instability mechanisms responsible for the generation of corrugation waves~\citep{adda-bediaDynamicStabilityCrack2013} and crack front segmentation~\citep{ponsHelicalCrackfrontInstability2010,Leblond.11,Chen.15}. However, for heterogeneous materials with order unity toughness contrasts, the linear theory is of little use. Understanding nonlinear perturbations is also needed to determine how heterogeneity affects energy dissipation in fracture, since 1\textsuperscript{st}-order perturbations have a net zero contribution to dissipation. The importance of nonlinear effects has been demonstrated in a recent work~\citep{kolvinNonlinearFocusingDynamic2017}. There, a nonlinear equation of motion for crack fronts was derived in the context of scalar elasticity and their response to externally induced perturbations was computed. It was suggested that nonlinear front focusing coupled with the rate-dependence of fracture energy dissipation may govern the transition to micro-branching. 
		
In the framework of LEFM, the modeling of the crack front dynamics necessitates the knowledge of the variation of dynamic SIF with the crack front geometry. The mathematical foundations of nonlinear perturbations of static crack fronts are well-developed~\citep{leblondSecondorderCoplanarPerturbation2012,vasoyaFiniteSizeGeometrical2016,Adda-Bedia.06}. While the general methods for dynamic cracks were developed by~\cite{movchanPerturbationsPlaneCracks1998},~\cite{Ramanathan.97,RamanathanThesis},~\cite{norrisMultiplescalesApproachCrackfront2007} and~\cite{willisCrackFrontPerturbations2013}, these works did not provide \textit{explicit, tractable} formulas that can be applied to concrete crack evolution. \cite{Morrissey.00} obtained an explicit time-dependent 1\textsuperscript{st}-order formula for the energy-release-rate of Mode~I crack fronts, that was used to numerically propagate crack fronts in heterogeneous solids. Based on these works, we compute the 2\textsuperscript{nd}-order corrections to the energy-release-rate in the scalar model of elasticity and conventional ``vectorial'' elastodynamics. The energy balance between the elastic energy flux into the crack front and the dissipation can then be used to obtain equations of motion for dynamic planar crack fronts. 

\begin{figure}[ht]
	\centering
	\includegraphics[scale=1]{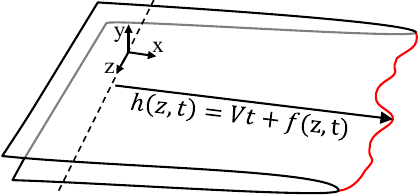}
	\caption{Geometry of the problem. The crack front (red) subtends two semi-infinite planar crack faces . The front propagates in the $y=0$ plane of a linear elastic solid at a velocity $V$ with a superimposed spatiotemporal perturbation $f(z,t)$.}\label{appendixC_problem_definition}
\end{figure}

 Consider an infinite body containing a semi-infinite planar crack that is driven by remote loading~(Fig.~\ref{appendixC_problem_definition}). The leading edge of the propagating crack front experiences perturbations in space and time around a steady motion at a constant velocity $V$. The coordinate system is defined as follows: $x$ is the crack propagation direction, the $y$ axis is perpendicular to the fracture $xz$ plane, and $t$ signifies time. The instantaneous crack front position is $h(z,t) = Vt + f(z,t)$, where $f$ is the crack front perturbation. Stresses vanish on the crack faces behind the crack front and diverge asymptotically as $(x-h(z,t))^{-1/2}$ ahead of the crack front where displacements are zero. To obtain expressions for the local energy-release-rate $G(z,t)$ at the front, we study two models: (i) scalar linear elasticity, which is analogous to antiplane shear, where a single wave equation determines a scalar displacement field. In this framework, the near-front fields are obtained by matching two asymptotic expansions as in~\cite{norrisMultiplescalesApproachCrackfront2007}. (ii) Tensile Mode~I fracture which is governed by the longitudinal and shear wave equations. There, we utilize a self-consistent expansion as in~\cite{RamanathanThesis}. These methods are readily generalized to higher-order perturbations.

In section~\ref{section:summary}, the explicit 2\textsuperscript{nd}-order formulae for the local energy-release-rates are summarized. We show that they reduce to known expressions for quasi-static fronts and 2D fracture. The detailed derivations are laid out for interested readers in \ref{scalarSection} for the scalar elastic model, and in \ref{vectorCracks} for Mode I fracture. In section~\ref{section:traveling_wave}, the  linear and 2\textsuperscript{nd}-order expressions are applied to the case of a traveling wave perturbation and the two models are compared. We end with a discussion of the results and their implication to predicting crack front dynamics.  

\section{Summary of the results}
\label{section:summary}

\subsection{Derivation of \textit{G} in the scalar elasticity model}

In \ref{scalarSection}, we solve for the asymptotic displacement and stress fields surrounding a crack in a model scalar elastic solid, extending the calculations of \cite{norrisMultiplescalesApproachCrackfront2007}. Fracture in scalar elasticity is identical to anti-plane shear (Mode III) fracture in a body constrained to displace along a single dimension, $\mathbf{u} = (0,0,u_z)$. The theory is characterized by a single wave speed ($b=1$). We assume cracks are driven at a velocity $V$ by remote loading conditions at a distance $l$ from the crack front. The elastic fields are expanded in the small parameter $\epsilon \equiv ||f||/l \ll 1$ in two separate perturbation series: the terms of the first series are centered at the perturbed crack front $x = h(z,t)$; the terms of the second series are centered at the unperturbed imaginary position of the crack front $x = Vt$. Solutions of the elasticity equation contain unknown integration constants determined by matching the two expansions. The main result of this calculation is the local energy-release-rate $G(z,t)$, exact up to $\order{f^2}$ and $\order{\epsilon^2}$. When $\epsilon\rightarrow 0$,
\begin{equation}\label{appenidx_eq_G_decomposed}
G(z,t) = G_r g(V_\perp)\left(1+H^{(1)}[f]+H^{(2)}[f,f] +\order{f^3}\right)\,,
\end{equation}
where the rest energy-release-rate $G_r$ is determined by the loading conditions, $g(V_\perp)$ is the dynamical contribution given by~\citep{riceThreedimensionalPerturbationSolution1994},
\begin{equation}\label{appenidx_eq_g(V)}
g(V_\perp) =\sqrt{\frac{1-V_\perp}{1+V_\perp}}\,,
\end{equation}
and $V_\perp(z,t) = (V+f_t)/\sqrt{1+f_z^2}$ is the local normal front velocity of the crack front. Subscripted functions $f_t,\,f_z$ denote partial derivatives. The linear functional $H^{(1)}[f]$ and the 2\textsuperscript{nd}-order functional $H^{(2)}[f,f]$  depend only on the crack front history $\lbrace f(z,t'<t)\rbrace$,
\begin{eqnarray}
H^{(1)}[f] &=& -\frac{1}{\alpha^2} \Psi[f]\,,\label{appendix_eq_h1}\\
H^{(2)}[f,f] &=&  \frac{1}{4\alpha^4}\Psi[f]^2 +\frac{1}{2\alpha^4}\Psi[f\Psi[f]]-\frac{1-2 V}{4\alpha^4}\Psi_2[f^2]-\frac{1+2 V}{2\alpha^4}f\Psi_2[f]\,,\label{appendix_eq_h2}
\end{eqnarray}
where $\alpha=\sqrt{1-V^2}$, and, with the definition of the Fourier transform $\bar{f}(k,t) = \int \!\mathrm{d}z\, e^{-ikz} f(z,t)$,
\begin{eqnarray}
\label{section2_Psi_eqs}
\overline{\Psi[f]}(k,t) &=& \alpha k \int_{-\infty}^t \!\mathrm{d}t'\,\frac{J_1(\alpha k (t-t'))}{t-t'}\bar{f}(k,t')\;,\\
\label{section2_Psi2_eqs}\overline{\Psi_2[f]}(k,t) &=& \alpha^2 k^2\int_{-\infty}^t\!\mathrm{d}t' \frac{J_2(\alpha k (t-t'))}{t-t'}\bar{f}(k,t')\;,
\end{eqnarray}
where $J_1$ and $J_2$ are the 1\textsuperscript{st}- and 2\textsuperscript{nd}-order Bessel functions.

\subsubsection{Recovery of known limits}\label{subsection_scalar_limits}

To test Eq.~(\ref{appenidx_eq_G_decomposed}) against known results, we consider two limiting cases: that of a straight front ($f_z=0$), and that of a steadily propagating front ($f_t=0$).T
he condition $f_z=0$ is equivalent to taking $k\rightarrow 0$ in Eqs.~(\ref{section2_Psi_eqs}-\ref{section2_Psi2_eqs}), which approach $\Psi[f],\Psi_2[f]\rightarrow 0$. Then the energy-release-rate, Eq.~(\ref{appenidx_eq_G_decomposed}), approaches $G\rightarrow G_r g(V+f_t)$, which is identical to the equation of motion of a 2D crack~\citep{riceThreedimensionalPerturbationSolution1994}. 

When $f_t =0$, Eqs.~(\ref{section2_Psi_eqs}-\ref{section2_Psi2_eqs}) become $\overline{\Psi[f]} = \alpha |k| \bar{f}(k)$ and $\overline{\Psi_2[f]} = \frac{1}{2}\alpha^2 k^2 \bar{f}(k)$. Then
\begin{eqnarray}
&&\overline{H^{(1)}[f]} = -\frac{1}{\alpha} |k|\bar{f}(k)\,,\nonumber\\
&&\overline{H^{(2)}[f,f]} = \\
&&\frac{1}{4\alpha^2}\int \!\mathrm{d}k'\, \bar{f}(k-k')\bar{f}(k') \left(|k'||k-k'| +|k|(|k'|+|k-k'|)-k^2+(1+2 V)k'(k-k')\right)\,.\nonumber
\end{eqnarray}
Using these results and expanding $g(V_\perp)\simeq g(V) -\frac{1}{2} Vg'(V)f_z^2$, 
Eq.~(\ref{appenidx_eq_G_decomposed})  becomes
\begin{equation}
\begin{split}\label{appendix_eq_g2_steady}
\overline{G}(k) &= G_r g(V)\Bigg\{ 2\pi\delta(k) - \frac{1}{\alpha} |k|\bar{f}(k) \\
&+\frac{1}{4\alpha^2}\int \!\mathrm{d}k'\, \bar{f}(k-k')\bar{f}(k') \left(|k'||k-k'| +|k|(|k'|+|k-k'|)-k^2+k'(k-k')\right)\Bigg\}\,,
\end{split}
\end{equation}
after symmetrization $k'\leftrightarrow k-k'$. In the limit $V\rightarrow 0$, Eq.~(\ref{appendix_eq_g2_steady}) becomes
\begin{equation}
\begin{split}\label{appendix_eq_g2_static}
\overline{G}(k) &= G_r \Bigg\{ 2\pi\delta(k) -  |k|\bar{f}(k) \\
&+\frac{1}{4}\int \!\mathrm{d}k'\, \bar{f}(k-k')\bar{f}(k') \left(|k'||k-k'| +|k|(|k'|+|k-k'|)-k^2+k'(k-k')\right)\Bigg\}\,.
\end{split}
\end{equation}
The same expression is obtained from the 2\textsuperscript{nd}-order approximation of the SIF of quasi-static crack fronts~\citep{leblondSecondorderCoplanarPerturbation2012,vasoyaFiniteSizeGeometrical2016}).

\subsection{Derivation of \textit{G} in Mode I fracture}\label{subsection_Mode_I_summary}
		
In \ref{vectorCracks}, we determine the asymptotic fields in the vicinity of a crack that propagates under remote tensile loading conditions in an infinite body of Young modulus $E$ and Poisson ratio $\nu$, extending the calculations of \cite{RamanathanThesis}. Without perturbations, the loading conditions produce an asymptotic Mode~I stress field. Ahead of the crack front at the fracture plane, the tensile stress is 
$$
\sigma_{yy} \sim K_0^\sigma/\sqrt{2\pi (x-Vt)}\,,
$$
where the SIF is $K_0^\sigma = K_r^\sigma k(V)$,  $K_r^\sigma$ is the rest SIF, and $k(V)$ is a universal function of the crack velocity defined in Eq. (6.4.26) of~\cite{Freund.90}. The energy-release-rate is then $G = G_r g(V)$ where the rest energy-release-rate is $G_r = \frac{1-\nu^2}{E} (K_r^\sigma)^2$. The dynamical prefactor is $g(V)=A_I(V) k(V)^2$ where $A_I(V) = (1-\nu)^{-1}(V^2\alpha_a)/(R(V) b^2)$;  $R(V) = 4\alpha_a\alpha_b - (1+\alpha_b^2)^2$;  $\alpha_s = \sqrt{1-(V/s)^2}$;  and $s=a,b$ are the longitudinal and the shear wave speeds~\citep{Freund.90}.

For space-time-dependent crack fronts, we derive a perturbation expansion of the energy-release-rate
\begin{equation}\label{eq_modeI_G_expansion}
    G(z,t) = G_r g(V_\perp)\left(1+H^{(1)}[f]+H^{(2)}[f,f] +\order{f^3}\right)
\end{equation}
Explicit expressions for the history functionals $H^{(1)}[f]$ and $H^{(2)}[f,f]$ are found in the space $(k,t)$, defined via the Fourier transform $\bar{f} = \int \!\mathrm{d}z \, e^{-ikz}f(z,t)$. The linear functional is $\overline{H^{(1)}[f]} = 2I_1[\bar{f}]$ where $I_1[\bar{f}] = -\int_{-\infty}^t\!\mathrm{d}t'\,A_1(k,t-t')\bar{f}(k,t')$ and~\citep{Morrissey.00}
\begin{equation}
    A_1(k,t) = k^2\left[-\frac{a}{2} \frac{J_1(\beta_a t)}{\beta_a t} + c \frac{J_1(\beta_c t)}{\beta_c t} - \frac{1}{4} \int_b^a\!\mathrm{d}\eta\,\Theta(\eta)\left(\frac{\eta^2+V^2}{\eta^2-V^2}J_2(\beta_\eta t)-J_0(\beta_\eta t)\right)\right]\,,
\end{equation}
where $\beta_s = \sqrt{s^2 - V^2} |k|$.
The 2\textsuperscript{nd}-order functional is 
\begin{equation}
\begin{split}
    \overline{H^{(2)}[f,f]} &= 2 I_1[\bar{f}*I_1[\bar{f}]]-\frac{1}{2}I_1[I_1[\bar{f}*\bar{f}]]-\bar{f}*I_1[I_1[\bar{f}]]+I_1[\bar{f}]*I_1[\bar{f}]\nonumber\\
        &+\frac{V}{2}I_2[\bar{f}*\bar{f}] - V \bar{f}*I_2[\bar{f}]\nonumber\,.
\end{split}
\end{equation}
where the convolution operator is defined as $f*g = (2\pi)^{-1}\int \!\mathrm{d}k'\,f(k-k')g(k')$, and $I_2[\bar{f}] = -\int_{-\infty}^t\!\mathrm{d}t' \, A_2(k,t-t')\bar{f}(k,t')$ with
$$
 A_2(k,t) = |k|^3\Bigg[-\gamma_a \frac{J_2(\beta_a t)}{\beta_a t} + 2\gamma_c \frac{J_2(\beta_c t)}{\beta_c t}-\frac{1}{4} \int_b^a\!\mathrm{d}\eta\,\frac{\Theta(\eta)}{\sqrt{\eta^2-V^2}}\left(\frac{3\eta^2+V^2}{\eta^2-V^2}J_3(\beta_\eta t)-J_1(\beta_\eta t)\right)\Bigg]\,.
$$
These expressions can be used to evolve crack fronts in time ~\citep{Morrissey.00}.

\begin{figure}[ht]
\centerline{\includegraphics[scale = 1]{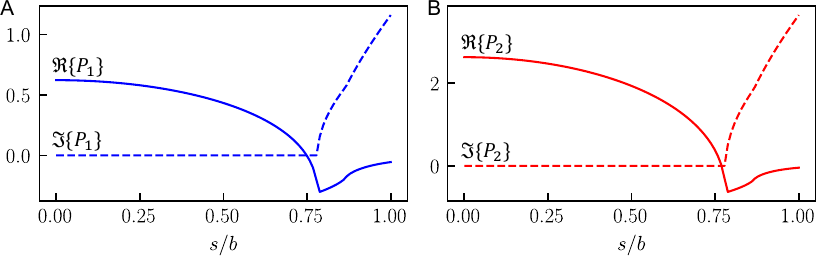}}
	\caption{(A) Real and imaginary parts of $P_1(s;V,\nu)$  (B) and  $P_2(s;V,\nu)$. $V=0.5 b$ and $\nu = 0.3$.}\label{fig2}
\end{figure}

\begin{figure}[ht]
\centerline{\includegraphics[scale = 1]{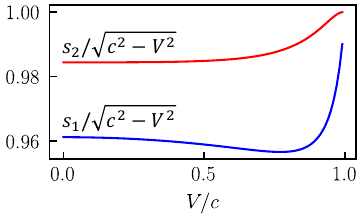}}
	\caption{The real roots $s_1$ of $P_1$ and $s_2$ of $P_2$ normalized by $\sqrt{c^2-V^2}$. $\nu = 0.3$.}\label{fig3}
\end{figure}

It is useful to write $G$ in wavenumber-frequency space $(k,\omega)$, where $\hat{f}(k,\omega) = \int\!\mathrm{d} z \mathrm{d} t \, e^{-i\omega t -ikz}f(z,t)$. We define the convolution $\hat{f}\otimes\hat{g} = (2\pi)^{-2}\int\!\mathrm{d}k'\mathrm{d}\omega'\,\hat{f}(k-k',\omega-\omega')\hat{f}(k',\omega')$. The perturbation expansion of the energy-release-rate is then
\begin{equation}\label{eq:Mode_I_G_Fourier}
\widehat{G}(k,\omega)= G_rg(V)\left((2\pi)^2\delta(k)\delta(\omega)+\widehat{\delta G^{(1)}}(k,\omega)+\widehat{\delta G^{(2)}}(k,\omega)+\order{f^3}\right)\,,
\end{equation}
where the 1\textsuperscript{st}-order correction is~\citep{Ramanathan.97}
\begin{equation}\label{delta_G1_fourier}
\widehat{\delta G^{(1)}}(k,\omega)=-2|k| P_1(\omega/|k|;V,\nu)\hat{f}(k,\omega)\,,
\end{equation}
and the 2\textsuperscript{nd}-order correction is
\begin{equation}\label{delta_G2_fourier}
\begin{split}
&\widehat{\delta G^{(2)}}(k,\omega)=
 2|k|P_1\left\{ \hat{f}\otimes\left(|k|P_1 \hat{f}\right)\right\} 
-\left\{ \frac{1}{2}k^2 P_1^2+\frac{i}{2}V\omega|k|P_2\right\}(\hat{f}\otimes \hat{f})\\& -\hat{f}\otimes\left\{ \left(k^2 P_1^2-iV\omega|k| P_2\right) \hat{f}\right\}   +(|k|P_1 \hat{f})\otimes(|k|P_1 \hat{f})\,.
\end{split}
\end{equation}
The kernel functions $P_1$ and $P_2$ can be computed explicitly by the formulae
\begin{align}\label{P1}
	P_1(s;V,\nu) = &-\frac{1}{2}\gamma_a\sqrt{1-\frac{\gamma_a^2 s^2}{a^2}}+\gamma_c\sqrt{1-\frac{\gamma_c^2 s^2}{c^2}}\nonumber\\
	&+\frac{1}{2}\int_{b}^{a}\frac{(s^2+V^2) (\eta^2+V^2)-2 \eta^2 V^2}{  \sqrt{\eta^2-(s^2+V^2)} \left(\eta^2-V^2\right)^2}  \Theta(\eta)\,\mathrm{d}\eta\,,
\end{align}
and
\begin{align}\label{P2}
	P_2(s;V,\nu) = &2\frac{\gamma_c^3}{c^2}\sqrt{1-\frac{\gamma_c^2s^2}{c^2}}-\frac{\gamma_a^3}{a^2}\sqrt{1-\frac{\gamma_a^2s^2}{a^2}}+\nonumber\\
	&\int_{b}^{a}\frac{(s^2+V^2) (3\eta^2+V^2)-2 \eta^2(\eta^2+ V^2)}{ \sqrt{\eta^2-(s^2+V^2)} \left(\eta^2-V^2\right)^3}  \Theta(\eta)\,\mathrm{d}\eta\,,
\end{align}
where $\gamma_s = 1/\sqrt{1-V^2/s^2}$, $c$ is the Rayleigh wave speed, and
\begin{equation}
\Theta(\eta) = \frac{2}{\pi}\arctan\left[4\sqrt{1-\frac{\eta^2}{a^2}}\sqrt{\frac{\eta^2}{b^2}-1}\bigg/\left(2-\frac{\eta^2}{b^2}\right)^2\right]\,.
\end{equation}
The branch cuts of the square roots in Eqs.~(\ref{P1},\ref{P2}) are defined such that $\sqrt{1-s^2}=i\,\mathrm{sign}(s)\sqrt{s^2-1}$ for $|s|>1$. 
Example curves of the functions $P_1$ and $P_2$ are depicted in Fig.~\ref{fig2}. Particularly, the zeros $s_1$ and $s_2$ of $P_1$ and $P_2$ do not coincide~(Fig.~\ref{fig3}). The roots $s = \pm s_1$ of $P_1$ signify the existence crack front waves, since $\delta G^{(1)} = 0$ at $\omega/|k| =\pm s_1$~\citep{Ramanathan.97,morrisseyCrackFrontWaves1998,Morrissey.00}. Because $s_1\neq s_2$, $\delta G^{(2)}\neq 0$ at $\omega/|k| =\pm s_1$. Therefore, crack front waves are expected to have a non-zero contribution to the energy-release-rate at the 2\textsuperscript{nd}-order.

\subsubsection{Recovery of known limits}
Eq.~(\ref{eq_modeI_G_expansion}) can be reduced to known expressions at the 2D limit $f = f(t)$ and at the quasi-static limit $f = f(z)$. 
In the 2D limit, $H^{(1)}[f] = 0$ and $H^{(2)}[f,f] = 0$ since $\bar{f}\propto \delta(k)$ and the 2D expression for the energy-release-rate $G = G_r g(V+f_t)$ is recovered.
In the quasi-static limit, $f = f(z)$. Then, $I_1[\bar{f}] = -P_1(0;V)|k|$ and $I_2[\bar{f}] = \pi_1 k^2$ where
\begin{equation}\label{section2_eq_pi1}
\pi_1 = \frac{1}{2}\frac{a}{a^2-V^2} -\frac{c}{c^2-V^2} +\frac{1}{2}\int_b^a \frac{\eta^2+V^2}{(\eta^2-V^2)^2}\Theta(\eta)\mathrm{d}\eta\,.
\end{equation}
The history functionals become 
$$
\overline{H^{(1)}[f]} = -2 P_1(0;V)|k|\bar{f}(k).
$$
and
\begin{equation}
\begin{split}
\overline{H^{(2)}[f]} =  &P_1(0;V)^2 \int\!\mathrm{d}k'\, \bar{f}(k-k')\bar{f}(k')\left(|k|(|k'|+|k-k'|) -k^2+k'(k-k') + |k'||k-k'|\right)\nonumber\\
&+V \pi_1 (k\bar{f})*(k\bar{f})\nonumber\,.
\end{split}    
\end{equation}
Since $2\pi_1 =  g'(V)/g(V)$, the energy-release-rate expansion becomes
\begin{equation}
\begin{split}
    \bar{G}(k) &= G_r g(V)\Bigg\{2\pi\delta(k) -2 P_1(0;V) |k|\bar{f}(k) \\
    &+ P_1(0;V)^2 \int\!\mathrm{d}k'\, \bar{f}(k-k')\bar{f}(k')\left(|k|(|k'|+|k-k'|) -k^2+k'(k-k') + |k'||k-k'|\right)+\order{f^3}\Bigg\}\,.
\end{split}
\end{equation}
At the limit $V\rightarrow 0$, Eq.~(\ref{appendix_eq_g2_static}) is reproduced.

\section{Application of the formulae to the case of a sinusoidal traveling wave}\label{section:traveling_wave}
To gain insight into the 2\textsuperscript{nd}-order corrections to $G$, we apply the formulae to the case of a unit amplitude wave that travels in the positive $z$ direction
$$
f(z,t) = \cos(2\pi( z - \varv t)/L)\,,
$$  
where $L>0$ is the wavelength and $\varv>0$ is the phase velocity.
\subsection{Scalar elasticity}

For $\varv<\alpha$, the corrections to the energy-release rate given by Eqs.~(\ref{appendix_eq_g1}-\ref{appendix_eq_g2}) are
\begin{eqnarray}\label{eq:scalar_sinusoidal}
	\delta G^{(1)} &=& -\frac{2\pi}{L}\alpha^{-2}\sqrt{\alpha^2-\varv^2}\cos(2\pi(z-\varv t)/L) \nonumber\\
	\delta G^{(2)} &=&\left(\frac{2\pi}{L}\right)^2\left\{ \frac{\alpha^2-\varv^2}{4\alpha^4}\cos(4\pi(z-\varv t)/L)
	-\frac{\sqrt{\alpha^2-\varv^2}V \varv}{2\alpha^4} \sin(4\pi(z-\varv t)/L)\right\}\,.
\end{eqnarray}

We compared the corrections $\delta G^{(1)}$ and $\delta G^{(1)}+ \delta G^{(2)}$ for $L= 2\pi$, $V = 0.5 b$ and $\varv = 0.5\alpha$~(Fig. \ref{fig4}A). The 2\textsuperscript{nd}-order contribution increased the energy-release-rate at the troughs of the traveling wave and decreased it at the crests, similarly to a static perturbation. However, the traveling wave also induced an out-of-phase contribution that was retarded relative to the wave. To examine the significance of the 2\textsuperscript{nd}-order correction, we computed the ratio of its norm, $|\delta G^{(2)}|$, to that of the 1\textsuperscript{st}-order correction, $|\delta G^{(1)}|$, where the norm is defined as $|f(z)| = \sqrt{\int\!\mathrm{d}z\,f(z)^2}$~(Fig. \ref{fig4}C). At low crack velocities, the relative magnitude $\delta G^{(2)}$  decreased with $\varv$ even as $\delta G^{(1)}\rightarrow 0 $ when $\varv\rightarrow \alpha$. Above $V \sim 0.5b$ the opposite trend was observed, where $\delta G^{(2)}$ became increasingly dominant at higher $V$ and $\varv$.
 
\subsection{Mode I fracture}

The choice of a sinusoidal function makes it straightforward to use Eqs.~(\ref{delta_G1_fourier}-\ref{delta_G2_fourier}).
Taking $\varv<c_R$ and Fourier transforming to real space we obtain,
\begin{eqnarray}
	\delta G^{(1)} &=& -\frac{4\pi}{L}P_1(\varv)\cos(2\pi(z-\varv t)/L)\\
	\delta G^{(2)} &=&\left(\frac{2\pi}{L}\right)^2\left\{P_1(\varv)^2\cos(4\pi(z-\varv t)/L)-\frac{1}{2}V \varv P_2(\varv)\sin(4\pi(z-\varv t)/L)\right\}\,.
\end{eqnarray}

\begin{figure}[ht]
	\centerline{\includegraphics[scale = 1]{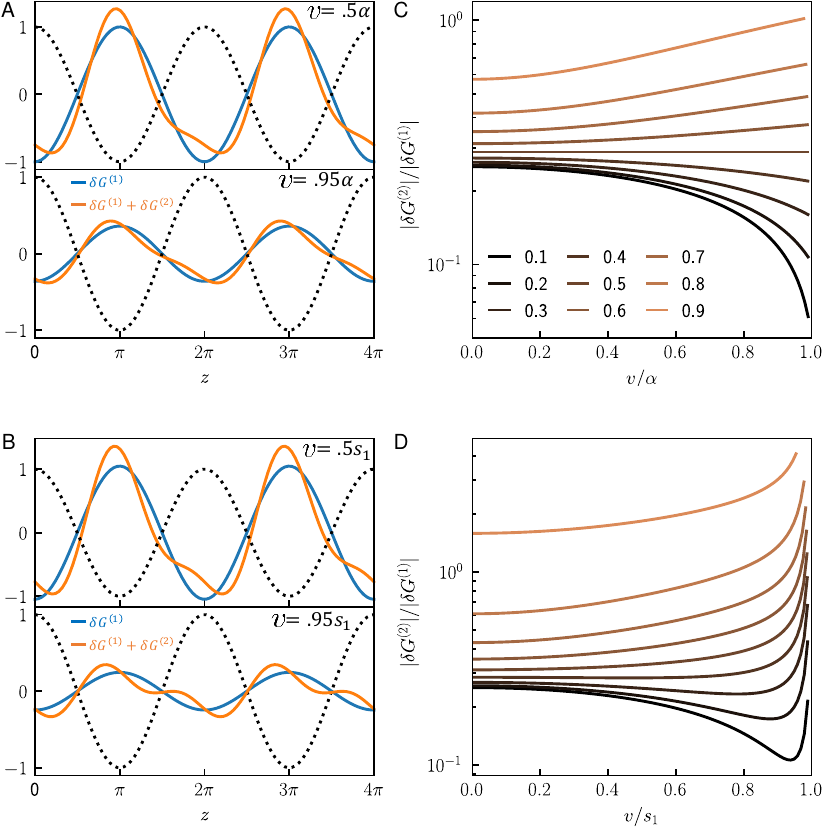}}
	\caption{ Energy-release-rate corrections for a traveling wave perturbation $f = \cos(z-\varv t)$ (black dotted line). (A) Scalar elasticity. $V=0.5 b$. (B) Mode I fracture. $\nu=0.3$, $V=0.5 b$. Ratios of the 2\textsuperscript{nd}-order correction to the 1\textsuperscript{nd}-order correction for (C) scalar elasticity and (D) Mode I fracture. Colors denote the crack velocity $V$.}\label{fig4}
\end{figure}

These expressions lend themselves to the following interpretation. Qualitatively similar to the energy-release-rate of a static sinusoidal front, at the 1\textsuperscript{st}-order, $G$ is lower at the crests and higher at the troughs of the traveling wave for $\varv<s_1$, where $s_1$ is the front wave speed~(Fig. \ref{fig4}B). The 2\textsuperscript{nd}-order correction has two contributions: an in-phase term that acts to increase the energy-release-rate at the crests and troughs and to decrease it at the nodes; and an out-of-phase term that skews the energy-release-rate distribution in the direction of the wave propagation at the crests and vice versa at the troughs. As in scalar elasticity, the out-of-phase contribution is retarded relative to the traveling wave.
In particular, when a disturbance travels at the front wave velocity, $\varv = s_1$,  the 2\textsuperscript{nd}-order out-of-phase term is the only contribution to the energy-release-rate. Thus, front waves are predicted to produce second harmonic variation of the energy-release-rate along the crack front that is strongest at the wave nodes.

We estimated the significance of the 2\textsuperscript{nd}-order correction by evaluating the ratio of its norm to that of the 1\textsuperscript{st}-order correction~(Fig. \ref{fig4}D). Similarly to scalar elasticity, the ratio $|\delta G^{(2)}|/|\delta G^{(1)}|$ monotonically decreased for $\varv\ll s_1$ and small $V$. This trend was reversed above $V\sim 0.5 b$. However, unlike the scalar case, $|\delta G^{(2)}|/|\delta G^{(1)}|$ diverged at $\varv=s_1$ for all crack velocities. 

\section{Discussion}
In the previous sections, we provided 2\textsuperscript{nd}-order perturbation expansions for the local energy-release-rate of planar crack fronts with spatiotemporally variable configurations in two theoretical frameworks: scalar elasticity and Mode I fracture. While more complex, the Mode I theory is a first-principles description that successfully explains tensile crack propagation~\citep{Sharon.99,goldmanAcquisitionInertiaMoving2010}. A comparison of the two theories helps elucidate the meaning and possible implications of the derived expressions. 

The two expansions have common features. They are both separable into products of a dynamical factor that depends on $V$ and of a historical part that convolves past crack front configurations with time-decaying kernels. In addition, both expansions become identical at the static limit $\omega\rightarrow 0,\, V\rightarrow 0$. Another point of similarity is that the 1\textsuperscript{st}-order corrections vanish at the dispersion curves which are $\omega =\pm \alpha k$ in the scalar elasticity case and $\omega =\pm s_1 k $ in the Mode I case. 

However, there is an important distinction between the two theories. In scalar elasticity, the 2\textsuperscript{nd}-order contribution is significant only for large crack velocities, since  $\delta G^{(2)}\sim\delta G^{(1)}\sim\mathcal{O}\left(\sqrt{\alpha^2 -\omega^2/k^2}\right)$  as $\omega \rightarrow \alpha k$. In Mode I fracture, the 1\textsuperscript{st}-order correction has a simple root at $\omega =\pm s_1 k$. The 2\textsuperscript{nd}-order correction, however, does not vanish there since $P_2(s_1)\neq P_1(s_1)=0$. Hence, at the crack front wave dispersion, the 2\textsuperscript{nd}-order terms become the leading correction in the expansion and provide a mechanism for wave-wave interactions along the crack front.

The 2\textsuperscript{nd}-order expansions for the energy-release-rate may be utilized to predict crack front dynamics through local energy balance $G(z,t)=\Gamma(z,x=h(z,t))$, where the local fracture energy $\Gamma$ is a material property and $x=h(z,t)$ is the instantaneous front configuration. Using the separable forms, Eqs.(\ref{appenidx_eq_G_decomposed},\ref{eq_modeI_G_expansion}), one obtains an equation of motion for the front normal velocity 
$$
V_\perp = g^{-1}\left[\frac{\Gamma}{G_r}(1-H^{(1)}[f]+H^{(1)}[f]^2-H^{(2)}[f,f])\right]\,,
$$
where $g^{-1}(\cdot)$ is the inverse function of $g(V)$. An equation of motion in the context of scalar elasticity was derived and numerically solved in \cite{kolvinNonlinearFocusingDynamic2017}, where 2\textsuperscript{nd}-order corrections resulted in focusing effects that were qualitatively similar to experimental observations of crack front dynamics during micro-branch formation. Alternatively, the dynamics can be resolved in Fourier space, for example using Eq.~(\ref{eq:Mode_I_G_Fourier}) for Mode I fracture, to compute the energy-release-rate part of energy balance~\citep{kolvinDualRoleHeterogeneity2024}. Future research will investigate how the nonlinear contributions to the energy-release-rate give rise to in-plane front roughness, and how wave-wave interactions modify energy dissipation at the crack front.

\appendix

\section{Nonlinear perturbation of scalar elastic crack fronts} \label{scalarSection}

This section develops a perturbation theory for planar crack fronts propagating in a solid described by a scalar displacement field $u$. The stress vector field is defined by $\mathbf{\sigma} = \mu\nabla u$ where $\mu$ is the elastic modulus. This formulation is identical to the elastic problem of anti-plane shear deformation when the orthogonal displacement components are artificially set to zero~\citep{riceThreedimensionalPerturbationSolution1994}. In general, freely propagating anti-plane shear (Mode~III) cracks experience tensile (Mode I) and in-plane shear (Mode II) stresses when perturbed~\citep{geubelleSpectralMethodThreedimensional1995}. The scalar model, however, is useful in obtaining a qualitative understanding of dynamic fracture, as it is much simpler than ``full'' elastodynamics~\citep{perrinDisorderingDynamicPlanar1994}. 

Displacements are governed by Newton's 2\textsuperscript{nd} law 
$$
\rho\partial_{t}^2u = \mu \nabla\cdot\mathbf{\sigma}
$$
where $\rho$ is the density. With the definition of the elastic wave velocity $b=\sqrt{\mu/\rho}$, the displacement satisfies the wave equation
\begin{equation}\label{appendix_eq_scalar_wave}
\nabla^2 u-\frac{1}{b^{2}}\partial_t^2 u=0\,.
\end{equation}
In the following, we set $b=1$ and
we assume that the crack is driven at velocity $V$ by stresses that are applied at a large distance $l$ from the crack front. It is helpful to transform the wave equation to the co-moving frame, $x\rightarrow(x-Vt)/l,\,y\rightarrow y/l,\, z\rightarrow z/l,\,t/l\rightarrow t,\,\sigma\rightarrow \sigma/\mu$,
such that
\begin{equation}\label{appendix_eq_wave_rescaled}
\alpha^2 u_{xx}+ u_{yy}+ u_{zz}- u_{tt}+2V u_{xt}=0\,,
\end{equation}
where subscripts denote partial derivatives.
The equation of motion is supplemented by the boundary conditions at the fracture plane $y=0$, 
\begin{equation}\label{appendix_eq_bc}
 u(x>f(z,t),y=0,z,t)=0;\;\; \sigma_y(x<f(z,t),y=0,z,t)= 0.
\end{equation}

For the simple crack, $f(z,t) = 0$, the asymptotic elastic fields that solve this problem are well-known~\citep{eshelbyElasticFieldCrack1969,riceThreedimensionalPerturbationSolution1994,norrisMultiplescalesApproachCrackfront2007}. The  displacement can be written as a Williams expansion
\begin{equation}\label{appendix_eq_simple_crack}
 u(x,y,z,t) =  \sqrt{\frac{2}{\pi}}\frac{K^u_0}{\mu}\,\Im\left\{ \sqrt{x-Vt +i \alpha y}\left[1+\frac{m}{3}(x-Vt +i \alpha y)+\frac{n}{5}(x-Vt +i \alpha y)^2+ ... \right]\right\}
\end{equation}
where $i=\sqrt{-1}$, $\alpha = \sqrt{1-V^2}$ and $ 
 K^u_0 =K^\sigma_0/\alpha$. The SIF $K^\sigma_0$ and the coefficients $m\propto l^{-1}$ and $n\propto l^{-2}$ are determined by the loading conditions. The SIF can be written as a product $K^\sigma_0=K_rk(V)$ where $K_r$ is the rest SIF and $k(V) = \sqrt{1-V}$~\citep{riceThreedimensionalPerturbationSolution1994}. The energy-release-rate for the straight crack is then $G_0= G_r g(V)$ where $G_r\propto K_r^2$ is the rest energy release and $g(V) = \sqrt{(1-V)/(1+V)}$.

The simple crack solution provides a basis for exploring the effect of arbitrary perturbations. 
Below, we follow the method of \cite{norrisMultiplescalesApproachCrackfront2007} where the elastic fields are determined by matching two asymptotic expansions. In the ``inner'' expansion the fields are centered at the crack front position $x=f(z,t)$. In the ``outer'' expansion, the fields are centered at $x=0$. The coefficients of the inner and outer solutions are then matched to yield the displacement and the stress intensity factors, defined as
\begin{eqnarray}
    K^u &=& \lim_{x-f(z,t)\rightarrow 0^-} (x-f(z,t))^{-1/2} \sqrt{\frac{\pi}{2}}\mu u(x,y=0,z,t)\\
    K^\sigma &=& \lim_{x-f(z,t)\rightarrow 0^+} \sqrt{2\pi(x-f(z,t))}\sigma_y(x,y=0,z,t)\,.
\end{eqnarray}
The energy release is then  $G\propto  K^u  K^\sigma$.
 
\subsection{The inner solution}

We define $\epsilon = \mathrm{max}_{z,t}|f(z,t)|/l$, and the ``inner'' variables $X = \epsilon x,Y=\epsilon y$. Writing the displacement field as a function of the inner variables $ U(X,Y,z;t) =  u(x,y,z;t)$, Eq.~(\ref{appendix_eq_wave_rescaled}) becomes
\begin{equation}\label{appendix_eq_inner}
\alpha^2 U_{XX}+ U_{YY}+2\epsilon V U_{Xt}+\epsilon^2( U_{zz}- U_{tt})=0\,.
\end{equation}
Expanding $ U  = \epsilon^{1/2} U^{(1/2)}+\epsilon^{3/2} U^{(3/2)}+\epsilon^{ 5/2} U^{(5/2)}+...$ and substituting the expansion in Eq.~(\ref{appendix_eq_inner}), $ U$ is found order-by-order in $\epsilon$. Since 2\textsuperscript{nd}-order corrections to $G$ are sought, we compute the expansion up to the $\epsilon^{ 5/2}$ term. The zeroth order solution corresponds to the dominant term in the simple crack field
\begin{equation}\label{appendix_eq_Phi12}
 U^{(1/2)} = \sqrt{\frac{2}{\pi}}\frac{K^u_0}{\mu} \Im{ S^{1/2}}\,,
\end{equation}
where $S = X - f +i\alpha Y$.

To find the next order terms, we will use the identity
\begin{equation}
(\alpha^2\partial_X^2+\partial_Y^2)w_1(\overline{S})w_2(S)=4\alpha^2
w_1'(\overline{S})w_2'(S)\,,
\end{equation} 
where $w_1(S),\,w_2(S)$ are arbitrary complex functions and $\overline{S}$ is the complex conjugate of $S$. The 1\textsuperscript{st}-order term in the expansion is then
\begin{equation}\label{appendix_eq_Phi32}
 U^{(3/2)} = \sqrt{\frac{2}{\pi}}\frac{K^u_0}{\mu} \Im{A^{(3/2)}S^{3/2}+B^{(3/2)} S^{1/2}+\frac{V}{4\alpha^2}f_t \overline{S}S^{-1/2}}\;.
\end{equation}
$A^{(3/2)}$ and $B^{(3/2)}$ are coefficients to be determined by matching to the outer solution. The 2\textsuperscript{nd}-order term in the expansion is 
\begin{equation}\label{appendix_eq_Phi52}
\begin{split}
& U^{(5/2)} = \\
&  \sqrt{\frac{2}{\pi}}\frac{K^u_0}{\mu} \Im\Bigg\{\left(A^{(5/2)}S^{5/2}+B^{(5/2)}S^{3/2}+C^{(5/2)}S^{1/2}\right)+ \left(\frac{mv}{2}f_t-vB^{(3/2)}_t+\frac{1}{2}f_{zz}-\frac{1}{2\alpha^2}f_{tt}\right)\frac{1}{2\alpha^2}\overline{S}S^{1/2}\\
&+\left(\frac{V}{2} B^{(3/2)}f_t+\frac{1+V^2}{4\alpha ^2}(f_t)^2-\frac{1}{4} (f_z)^2\right)\frac{1}{2\alpha ^2}\overline{S}S^{-1/2}-\frac{V^2}{16\alpha ^4}f_{tt}\overline{S}^2 S^{-1/2}-\frac{V^2}{32\alpha ^4}(f_t)^2\overline{S}^2 S^{-3/2}\Bigg\}\;.
\end{split}
\end{equation}
$A^{(5/2)}$, $B^{(5/2)}$ and $C^{(5/2)}$ are additional coefficients to be determined in the matching.

To make the matching to the outer solution straightforward, we expand Eqs.~(\ref{appendix_eq_Phi12},\ref{appendix_eq_Phi32},\ref{appendix_eq_Phi52}) around the unperturbed crack front position. Replacing $S=\epsilon^{-1}(s-\epsilon f)$ where $s=x+i\alpha y$, and gathering terms of the same order in $\epsilon$ we find the following expressions.
\newline

Terms of the order $\epsilon^0$:
\begin{equation}
 \sqrt{\frac{2}{\pi}}\frac{K^u_0}{\mu} \Im{s^{1/2}+\frac{1}{3} m s^{3/2}+A^{(5/2)}s^{5/2}}\;.
\end{equation}
Terms of the order $\epsilon^1$:
\begin{equation}
\begin{split}
& \sqrt{\frac{2}{\pi}}\frac{K^u_0}{\mu} \Im\Bigg\{-\frac{1}{2}f s^{-1/2} +(B^{(3/2)}-\frac{1}{2} m  f)s^{1/2}+\frac{V  f_t}{4\alpha^2}\overline{s}s^{-1/2}  +(B^{(5/2)}-\frac{5}{2}  A^{(5/2)} f)s^{3/2}\\
&-\frac{V^2  f_{tt}}{16\alpha^4}\overline{s}^2s^{-1/2}+\frac{1}{4}\left(\frac{m V}{\alpha^2}  f_t-\frac{2 V}{\alpha^2} B^{(3/2)}_t+\frac{1}{\alpha^2} f_{zz}-\frac{1}{\alpha^4}f_{tt} \right)\overline{s} s^{1/2}\Bigg\}\;.
\end{split}
\end{equation}
Terms of the order $\epsilon^2$:
\begin{equation}
\begin{split}
& \sqrt{\frac{2}{\pi}}\frac{K^u_0}{\mu} \Im \Bigg\{
-\frac{f^2  }{8 }s^{-3/2}+ \left(\frac{m}{8} f^2-\frac{f  B^{(3/2)}}{2}-\frac{V }{4\alpha^2} f f_t\right)s^{-1/2}+\frac{V  ff_t }{8\alpha^2}\overline{s}s^{-3/2} \\
&\left( C^{(5/2)}-\frac{3}{2} B^{(5/2)} f+\frac{15}{8} A^{(5/2)} f^2+\frac{ V}{2 \alpha ^2} B_t^{(3/2)}f-\frac{m V}{4 \alpha ^2}  f f_t+\frac{f f_{tt}}{4 \alpha ^4}-\frac{1}{4 \alpha ^2} f f_{zz}\right) s^{1/2}\\
&+\left(\frac{d}{dt}\left(-\frac{1}{16 \alpha ^2} m V f^2+\frac{1}{4 \alpha ^2} V B^{(3/2)} f+\frac{\left(1+V^2\right)}{16\alpha ^4}\frac{d}{dt}f^2\right)-\frac{1}{16 \alpha ^2} \frac{d^2}{dz^2}f^2\right)\overline{s}s^{-1/2}\\
&-\frac{V^2 }{64 \alpha ^4}\frac{d^2}{dt^2}f^2\overline{s}^2 s^{-3/2}\Bigg\}\;.
\end{split}
\end{equation}

The matching procedure is further simplified by introducing the polar coordinates $s=r e^{i\theta}$ and taking the imaginary part in the above  expressions:\\
Terms of the order $\epsilon^0$:
\begin{equation}\label{appendix_eq_inner_eps0_polar}
 \sqrt{\frac{2}{\pi}}\frac{K^u_0}{\mu} \left(\sqrt{r} \sin{\frac{\theta}{2}} +A^{(3/2)} r^{3/2}\sin\frac{3 \theta }{2}+ A^{(5/2)}r^{5/2} \sin\frac{5 \theta }{2} \right)\;.
\end{equation}
Terms of the order $\epsilon^1$:
\begin{equation}\label{appendix_eq_inner_eps1_polar}
\begin{split}
& \sqrt{\frac{2}{\pi}}\frac{K^u_0}{\mu}\Bigg\{ \frac{f }{2}r^{-1/2}\sin{\frac{\theta}{2}}+ \left(B^{(3/2)}-\frac{3}{2}A^{(3/2)} f\right) r^{1/2}\sin{\frac{\theta}{2}}-\frac{V  f_t}{4 \alpha ^2}r^{1/2}\sin\frac{3 \theta }{2}\\
& + \left(\frac{V B^{(3/2)}_t}{2 \alpha ^2}-\frac{3A^{(3/2)} V f_t}{4 \alpha ^2}+\frac{f_{tt}}{4 \alpha ^4}-\frac{f_{zz}}{4 \alpha ^2}\right)r^{3/2}\sin{\frac{\theta}{2}}\\ &+\left(B^{(5/2)}-\frac{5}{2} A^{(5/2)} f\right)r^{3/2} \sin\frac{3 \theta }{2}+\frac{V^2  f_{tt}}{16 \alpha ^4}r^{3/2}\sin\frac{5 \theta }{2}\Bigg\}\;.
\end{split}
\end{equation}
Terms of the order $\epsilon^2$:
\begin{equation}\label{appendix_eq_inner_eps2_polar}
\begin{split}
& \sqrt{\frac{2}{\pi}}\frac{K^u_0}{\mu}\Bigg\{\frac{f^2}{8 } r^{-3/2}\sin\frac{3 \theta }{2}-\frac{ V f f_t}{8 \alpha ^2}r^{-1/2}\sin\frac{5 \theta }{2}+ \left(\frac{1}{2} B^{(3/2)} f-\frac{ 3A^{(3/2)}f^2}{8}+\frac{ V f f_t}{4 \alpha ^2}\right)r^{-1/2}\sin{\frac{\theta}{2}}\\
&+  \left(\frac{15}{8} f^2 A^{(5/2)}-\frac{3}{2} f B^{(5/2)}+C^{(5/2)}+\frac{ V f B_t^{(3/2)}}{2 \alpha ^2}-\frac{ 3A^{(3/2)} V f f_t}{4 \alpha ^2}+\frac{f f_{tt}}{4 \alpha ^4}-\frac{f f_{zz}}{4 \alpha ^2}\right)r^{1/2}\sin{\frac{\theta}{2}}\\
&+\left(-\frac{ V f B_t^{(3/2)}}{4 \alpha ^2}-\frac{ V f_t B^{(3/2)}}{4 \alpha ^2}+\frac{ 3A^{(3/2)} V ff_t}{8 \alpha ^2}-\frac{\left(1+V^2\right) \left(f_t^2+f f_{tt}\right)}{8 \alpha ^4}+\frac{f f_{zz}}{8 \alpha ^2}+\frac{f_z^2}{8 \alpha ^2}\right)r^{1/2}\sin\frac{3 \theta }{2}\\ 
&+\frac{V^2 \left(f_t^2+f f_{tt}\right)}{32 \alpha ^4}r^{1/2} \sin\frac{7 \theta }{2}\Bigg\}\;.
\end{split}
\end{equation}

\subsection{The outer solution}

We derive an asymptotic solution for the outer problem, where the fields are expanded around the position of the unperturbed crack front. Let us consider an expansion involving a single Fourier component $ u =  u^{(0)}+\epsilon u^{(1)}+\epsilon^2 u^{(2)}+... =  u^{(0)}+ \sqrt{\frac{2}{\pi}}\frac{K^u_0}{\mu} \Im\{(\epsilon q^{(1)}(x,y)+\epsilon^2 q^{(2)}(x,y))e^{i(kz+\omega t)}\}$, where superscripts mark the order of each term in $\epsilon$. Here 
\begin{equation}\label{appendix_eq_outer_q0}
 u^{(0)} = \sqrt{\frac{2}{\pi}}\frac{K^u_0}{\mu} \Im\left\{s^{1/2}+\frac{m}{3}s^{3/2}+\frac{n}{5}s^{5/2}\right\}\;,
\end{equation} 
where $s,m$, and $n$ are the same as in the previous sections. The expansion is substituted in Eq.~(\ref{appendix_eq_wave_rescaled}), which is then solved term-by-term. Once a solution is found, the fields due to arbitrary perturbations can be computed by superposing Fourier components. To affect the matching of the inner and outer solutions, explicit expressions of $q^{(i)}$ will be equated with expressions~(\ref{appendix_eq_inner_eps0_polar}-\ref{appendix_eq_inner_eps2_polar}). Each $q^{(i)}(x,y)$ satisfies the equation
\begin{equation}\label{appendix_eq_q}
\alpha^2q_{xx}+q_{yy}+(\omega^2-k^2)q+2iV\omega q_{x}=0\;.
\end{equation}
A general solution to this equation is 
\begin{equation}\label{appendix_eq_q_solution}
q = \frac{1}{2\pi} \int_{-\infty}^\infty \!\mathrm{d}\xi\, \hat{q}(\xi)e^{i\omega\xi x+\omega\gamma(\xi) y}\;,
\end{equation}
where $\gamma^2 = \alpha^2\xi^2 +2V\xi + (k/\omega)^2-1$. Without loss of generality, we assume that $\omega>0$. To ensure the convergence of the integral~(\ref{appendix_eq_q_solution}), we must take the branch of $\gamma$ which has a positive real part. $\gamma$ has two branch points, $\xi = -\lambda_+$ and $\xi = \lambda_-$ where
\begin{equation}\label{appendix_eq_lambdapm}
\lambda_\pm = \frac{1}{\alpha ^2}\left(1-\alpha ^2k^2/\omega ^2\right)^{1/2}\pm \frac{V}{\alpha ^2}\,.
\end{equation}
It is convenient, then, to take one branch cut extending from  $\xi = -\lambda_+$ to $-i\infty$ and the other from  $\xi = \lambda_-$ to $+i\infty$. As we shall see, the exact shape of the branch cuts does not affect our calculations.

To determine $\hat{q}(\xi)$, we employ the boundary conditions Eq.~(\ref{appendix_eq_bc}). In terms of $q$ these boundary conditions are
\begin{equation}\label{appendix_eq_q_bc}
\begin{split}
q(x,0) &= 0; \; x>0\,,\\
q_y(x,0) &= 0;\; x<0\,.
\end{split}
\end{equation}
Fourier transforming the boundary conditions results in the relations
\begin{equation}\label{appendix_eq_q_analytical}
\begin{split}
\hat{q}(\xi)  &= \int_{-\infty}^0 \!\mathrm{d}x\, q(x,0) e^{-i\omega\xi x}\,, \\
-\omega\gamma(\xi)\hat{q}(\xi) &=\int_0^\infty \!\mathrm{d}x\, q_y(x,0) e^{-i\omega\xi x}\,.
\end{split}
\end{equation} 
These relations provide us with information about the analytical domain of $\hat{q}(\xi)$ and $\gamma(\xi)\hat{q}(\xi)$. The function $\hat{q}(\xi)$ is analytic for $\Im{\xi}<0$ and the function $\gamma(\xi)\hat{q}(\xi)$ is analytic for $\Im{\xi}>0$. We decompose $\gamma = \gamma^+\gamma^-$, where $\gamma^\pm =\alpha^{1/2} (\xi\pm\lambda_\pm)^{1/2}$ and $\gamma^+$ ($\gamma^-$) is analytic for $\Im{\xi}>0$ ($\Im{\xi}<0$), according to our choice of the branch of $\gamma$. Then, we readily see that $W \equiv \gamma^-\hat{q} = \gamma\hat{q}/\gamma^+$ is an entire function represented as a power series in $\xi$. In fact,  $\gamma^-q^{(n)}(x,y)$ should be a polynomial of order $n-1$. This can be understood from the previous section, where the solution in the inner variables becomes progressively more singular with increasing order. As $\xi$ is the Fourier conjugate of $x$, stronger singularities of $x$ translate into higher powers of $\xi$.

Let us consider the 1\textsuperscript{st}-order term $q^{(1)}(x,y)$, and set $\hat{q}^{(1)}(\xi) = q_0/\gamma^-$. For convenience, we write the $x,y$ coordinates as $\Re{s},\Im{s}$ respectively, and expand the integral in Eq.~(\ref{appendix_eq_q_solution}) in powers of $1/\xi$ (note that $\order{s\xi}\sim\order{1}$)
\begin{equation}
\begin{split}
&q^{(1)}(x,y) = \frac{1}{2\pi} \int \!\mathrm{d}\xi\, \frac{q_0}{\alpha^{1/2}(\xi-\lambda_-)^{1/2}}e^{i\omega\xi \Re{s}+\omega\gamma \Im{s}}=\\
& \frac{q_0}{2\pi\alpha^{1/2}} \int \!\mathrm{d}\xi\, e^{\omega  |\xi| \Im{s}+i \xi  \omega  \Re{s}}  \left(\frac{1}{\sqrt{\xi }}+\frac{\lambda_- -\omega  |\xi| \Im{s} \left(\lambda_- -\lambda _+\right)}{2 \xi^{3/2}}\right.\\
&\left. +\frac{3 \lambda _-^2+\xi^2 \omega ^2 \Im{s}^2 \left(\lambda _- -\lambda _+\right){}^2-\omega  |\xi| \Im{s} \left(3 \lambda _-^2+\lambda _+^2\right)}{8 \xi ^{5/2}}+\order{\xi^{-7/2}}\right)\;.
\end{split}
\end{equation}
We integrate this expression using the identity 
\begin{equation}
\begin{split}
\frac{1}{2\pi} \int \!\mathrm{d}\xi\, &\frac{e^{\omega  |\xi| \Im{s}+i \xi  \omega  \Re{s}}}{\xi^{n+1/2}}[1,\mathrm{sgn}\xi] = \\
&-\frac{2^{2n} n!}{(2n)!}\frac{e^{i\pi/4}}{\sqrt{\pi}}(-i)^n|\omega s|^{n-1/2}[\sin(n-\frac{1}{2})\theta,i\cos(n-\frac{1}{2})\theta]\;,
\end{split}
\end{equation}
where $s= r e^{i\theta}$ and $n\geq 0$.  A similar identity holds for the transformation $n\rightarrow -n$,
\begin{equation}
\begin{split}
\frac{1}{2\pi} \int \!\mathrm{d}\xi\, &\frac{e^{\omega  |\xi| \Im{s}+i \xi  \omega  \Re{s}}}{\xi^{-n+1/2}}[1,\mathrm{sgn}\xi] =\\ &\frac{(2n)!}{2^{2n} n!}\frac{e^{i\pi/4}}{\sqrt{\pi}}i^{-n}|\omega s|^{-n-1/2}[\sin(n-\frac{1}{2})\theta,-i\cos(n-\frac{1}{2})\theta]\,.
\end{split}
\end{equation}
Integrating and substituting $s=re^{i\theta}$, the 1\textsuperscript{st}-order solution is
\begin{equation}
\begin{split}
& q^{(1)}(x,y) =\frac{q_0 e^{i\pi/4}}{\sqrt{-\pi\omega } \alpha^{1/2}} \left[\frac{\sin\frac{\theta}{2}}{\sqrt{r}}+\sqrt{r} \left(-\frac{1}{4} i \omega  \sin\frac{\theta}{2} \left(3 \lambda _-+\lambda _+\right)+\frac{1}{4} i \omega  \sin\frac{3\theta}{2} \left(\lambda _--\lambda _+\right)\right)\right.\\
& +r^{3/2} \left(-\frac{1}{16} \omega ^2 \sin\frac{\theta}{2} \left(5 \lambda _-^2+2 \lambda _- \lambda _++\lambda _+^2\right)+\frac{1}{32} \omega ^2 \sin\frac{3\theta}{2} \left(5 \lambda _-^2-2 \lambda _- \lambda _+-3 \lambda _+^2\right)\right.\\
&\left.\left.-\frac{1}{32} \omega ^2 \sin\frac{5\theta}{2} \left(\lambda _--\lambda _+\right){}^2 \right)\right] \;.
\end{split}
\end{equation}
Substituting the explicit expressions~(\ref{appendix_eq_lambdapm}) for $\lambda_\pm$, the 1\textsuperscript{st}-order solution becomes 
\begin{equation}\label{appendix_eq_q1}
\begin{split}
& q^{(1)}(x,y) =\frac{q_0 e^{i\pi/4}}{\sqrt{-\pi\omega } \alpha^{1/2}} \left[r^{-1/2}\sin\frac{\theta}{2}+\left(\frac{i V \omega }{2 \alpha ^2}+\widehat{\Pi}\right) r^{1/2}\sin\frac{\theta}{2}-\frac{i V \omega  }{2 \alpha ^2}r^{1/2}\sin\frac{3\theta}{2}\right.\\
&\left. + \left(\frac{k^2}{2 \alpha ^2}-\frac{\omega ^2}{2 \alpha ^4}-\frac{V^2 \omega ^2}{4 \alpha ^4}+\frac{iV\omega \widehat{\Pi} }{2 \alpha ^2}\right) r^{3/2}\sin\frac{\theta}{2}\right.\\
&\left.+\left(\frac{V^2 \omega ^2}{8 \alpha ^4}-\frac{iV\omega \widehat{\Pi} }{2 \alpha ^2}\right) r^{3/2}\sin\frac{3\theta}{2}-\frac{V^2 \omega ^2 }{8 \alpha ^4}r^{3/2}\sin\frac{5\theta}{2}\right]\,. 
\end{split}
\end{equation}
Here, we have introduced a new symbol $\widehat{\Pi}$ with the definition
\begin{equation}\label{appendix_eq_Pi}
\begin{split}
\widehat{\Pi}(k,\omega) &= -i\alpha^{-2}\mbox{sign}(\omega)\sqrt{\omega^2-\alpha^2 k^2};\; \omega^2 >\alpha^2 k^2\,,\\ 
\widehat{\Pi}(k,\omega)  &= -\alpha^{-2}\sqrt{\alpha^2 k^2-\omega^2};\; \omega^2 <\alpha^2 k^2\,.
\end{split}
\end{equation} 
The 2\textsuperscript{nd}-order field $q^{(2)}(x,y)$ is similarly found by writing
$\hat{q}(\xi) = (q_1 +q_2\xi)/\gamma^-$, so that 
\begin{equation}
q^{(2)}(x,y) = \frac{1}{2\pi} \int \!\mathrm{d}\xi\, \frac{q_1 +q_2\xi}{\alpha^{1/2}(\xi-\lambda_-)^{1/2}}e^{i\omega\xi \Re{s}+\omega\gamma \Im{s}}\,.
\end{equation}
This integral can be separated into two parts, proportional to $q_1$ and $q_2$ respectively. We have already calculated the first part in Eq.~(\ref{appendix_eq_q1}). The order of the $r^{3/2}$ terms is higher than 2\textsuperscript{nd}-order since the 2\textsuperscript{nd}-order inner solution in Eq.~(\ref{appendix_eq_inner_eps2_polar}) does not contain them.

It remains therefore to calculate
\begin{equation}
\begin{split}
&\frac{1}{2\pi} \int \!\mathrm{d}\xi\, \frac{q_2\xi}{\alpha^{1/2}(\xi-\lambda_-)^{1/2}}e^{i\omega\xi \Re{s}+\omega\gamma \Im{s}}=\\
& \frac{q_2}{2\pi\alpha^{1/2}} \int \!\mathrm{d}\xi\, e^{\omega  |\xi| \Im{s}+i \xi  \omega  \Re{s}}\left(\sqrt{\xi }+\frac{\lambda _--\omega  | \xi | \Im{s} \left(\lambda _--\lambda _+\right)}{2 \sqrt{\xi }}\right.\\
&\left.+\frac{3 \lambda _-^2+\omega ^2 | \xi |^2 \Im{s}^2 \left(\lambda _--\lambda _+\right){}^2-\omega  | \xi | \Im{s} \left(3 \lambda _-^2+\lambda _+^2\right)}{8 \xi^{3/2}}+\order{\xi^{-5/2}}\right)\,.
\end{split}
\end{equation}
This is done in the same way as the 1\textsuperscript{st}-order and the result in polar coordinates is 
\begin{equation}
\begin{split}
&-\frac{q_2\; i e^{i\pi/4}}{2\sqrt{\pi } \alpha^{1/2}  \;(-\omega) ^{3/2}} \left[r^{-3/2}\sin\frac{3 \theta }{2}+\left(\frac{3 \, i\, V \;\omega }{2 \alpha ^2}+\widehat{\Pi}\right) r^{-1/2}\sin{\frac{\theta}{2}}-\frac{i V \omega  }{2 \alpha ^2}r^{-1/2}\sin\frac{5 \theta }{2}\right.\\
&\left.+\left(\frac{k^2}{\alpha ^2}-\frac{\omega ^2}{\alpha ^4}-\frac{7 V^2 \omega ^2}{8 \alpha ^4}+\frac{5 i V \omega \widehat{\Pi}}{2 \alpha ^2}\right) r^{1/2} \sin{\frac{\theta}{2}}+\left(-\frac{k^2}{2 \alpha ^2}+\frac{\omega ^2}{2 \alpha ^4}+\frac{3 V^2 \omega ^2}{4 \alpha ^4}-\frac{iV\omega \widehat{\Pi}}{2 \alpha ^2}\right) r^{1/2} \sin\frac{3 \theta }{2}\right.\\
&\left.-\frac{V^2 \omega ^2 }{8 \alpha ^4}r^{1/2}\sin\frac{7\theta}{2}\right]\,.
\end{split}
\end{equation}
The expression for the outer 2\textsuperscript{nd}-order field becomes
\begin{equation}\label{appendix_eq_q2}
\begin{split}
&q^{(2)}(x,y)=-\frac{q_2\; i e^{i\pi/4}}{2\sqrt{\pi } \alpha^{1/2}  \;(-\omega) ^{3/2}} \left[r^{-3/2}\sin\frac{3 \theta }{2}+\left(\frac{3 \, i\, V \;\omega }{2 \alpha ^2}+\widehat{\Pi}\right) r^{-1/2}\sin{\frac{\theta}{2}}-\frac{i V \omega }{2 \alpha ^2}r^{-1/2} \sin\frac{5 \theta }{2}\right.\\
&\left.+\left(\frac{k^2}{\alpha ^2}-\frac{\omega ^2}{\alpha ^4}-\frac{7 V^2 \omega ^2}{8 \alpha ^4}+\frac{5 i V \omega \widehat{\Pi}}{2 \alpha ^2}\right) r^{1/2}\sin{\frac{\theta}{2}}+\left(-\frac{k^2}{2 \alpha ^2}+\frac{\omega ^2}{2 \alpha ^4}+\frac{3 V^2 \omega ^2}{4 \alpha ^4}-\frac{iV\omega \widehat{\Pi}}{2 \alpha ^2}\right) r^{1/2}\sin\frac{3 \theta }{2}\right.\\
&\left.-\frac{V^2 \omega ^2 }{8 \alpha ^4}r^{1/2}\sin\frac{7\theta}{2}\right]\\
&+\frac{q_1 e^{i\pi/4}}{\sqrt{-\pi\omega } \alpha^{1/2}} \left[r^{-1/2}\sin\frac{\theta}{2}+ \left(\frac{i V \omega }{2 \alpha ^2}+\widehat{\Pi}\right) r^{1/2}\sin\frac{\theta}{2}-\frac{i V \omega  }{2 \alpha ^2}r^{1/2}\sin\frac{3\theta}{2}\right]\,.
\end{split}
\end{equation}

\subsection{Matching the solutions}

In the two previous sections, we have derived the asymptotic solutions until the 2\textsuperscript{nd}-order in $\epsilon$ for the inner fields Eqs.~(\ref{appendix_eq_inner_eps0_polar}-\ref{appendix_eq_inner_eps2_polar}) and for the outer fields Eqs.~(\ref{appendix_eq_outer_q0},\ref{appendix_eq_q1},\ref{appendix_eq_q2}). The expressions we have derived contain five inner coefficients $A^{(3/2)},A^{(5/2)},B^{(3/2)},B^{(5/2)},C^{(5/2)}$ and three outer coefficients $q_0,q_1,q_2$, which will be found by matching the inner and outer solutions. We note that, in general, the coefficients may be functions of $(z,t)$.

To match the terms in the two solutions we equate, order-by-order the prefactors of the polar functions $r^{-j/2}\sin( j\theta/2)$, where $j$ is an integer. The matching is made less straightforward since the inner solution is expressed in the real space coordinates $(z,t)$ and the outer solution is expressed in the Fourier coordinates $(k,\omega)$. However, we will stick to the present notation, with the implicit understanding that the outer solution terms must be first Fourier transformed before matching to the corresponding inner solution terms. We also adopt the notation $\Pi[f](z,t) = \frac{1}{(2\pi)^2}\int \!\mathrm{d}\omega\mathrm{ d}k\, e^{i(kz+\omega t)}\widehat{\Pi}(\omega,k)f(\omega,k)$. The final expressions will always be in the $(z,t)$ space.

\begin{table}[ht]
\begin{tabular}{c|c|c}
Term & Inner solution & Outer solution\\ \hline\hline
$\sqrt{r} \sin{\frac{\theta}{2}}$ & $ K^u_0$ & $ K^u_0$\\  \hline
$r^{3/2}\sin\frac{3 \theta }{2}$ & $A^{(3/2)}$ & $\frac{m}{3}$ \\ \hline
$r^{5/2} \sin\frac{5 \theta }{2}$ & $A^{(5/2)}$ & $\frac{n}{5}$
\end{tabular}
\caption{Terms of order $\epsilon^0$}\label{appendix_table_eps0}
\end{table}
\begin{table}[ht]
\begin{tabular}{c|c|c}
Term & Inner solution & Outer solution\\ \hline\hline
$r^{-1/2}\sin{\frac{\theta}{2}}$ & $\frac{f }{2}$& $\frac{q_0 e^{i\pi/4}}{\sqrt{-\pi\omega } \alpha^{1/2}}$ \\\hline
$r^{1/2}\sin{\frac{\theta}{2}}$ & $B^{(3/2)}-\frac{3}{2}A^{(3/2)} f$ & $\frac{q_0 e^{i\pi/4}}{\sqrt{-\pi\omega } \alpha^{1/2}}\left(\frac{i V \omega }{2 \alpha ^2}+\widehat{\Pi}\right)$\\\hline
$r^{1/2}\sin\frac{3 \theta }{2}$ & $-\frac{V  f_t}{4 \alpha ^2}$& $-\frac{q_0 e^{i\pi/4}}{\sqrt{-\pi\omega } \alpha^{1/2}}\frac{i V \omega  }{2 \alpha ^2}$ \\\hline
$r^{3/2}\sin{\frac{\theta}{2}} $ & $B^{(3/2)}_t\frac{V }{2 \alpha ^2}-A^{(3/2)}\frac{3 V f_t}{4 \alpha ^2}+\frac{f_{tt}}{4 \alpha ^4}-\frac{f_{zz}}{4 \alpha ^2}$ & $\frac{q_0 e^{i\pi/4}}{\sqrt{-\pi\omega } \alpha^{1/2}}\left(\frac{k^2}{2 \alpha ^2}-\frac{\omega ^2}{2 \alpha ^4}-\frac{V^2 \omega ^2}{4 \alpha ^4}+\frac{iV\omega \widehat{\Pi} }{2 \alpha ^2}\right)$\\\hline
$r^{3/2}\sin{\frac{3\theta}{2}}$& $B^{(5/2)}-\frac{5}{2} A^{(5/2)} f$ &$\frac{q_0 e^{i\pi/4}}{\sqrt{-\pi\omega } \alpha^{1/2}}\left(\frac{V^2 \omega ^2}{8 \alpha ^4}-\frac{iV\omega \widehat{\Pi} }{2 \alpha ^2}\right)$ \\\hline
$r^{3/2}\sin\frac{5 \theta }{2}$ & $\frac{V^2  f_{tt}}{16 \alpha ^4}$ &$-\frac{q_0 e^{i\pi/4}}{\sqrt{-\pi\omega } \alpha^{1/2}}\frac{V^2 \omega ^2 }{8 \alpha ^4}$
\end{tabular}
\caption{Terms of order $\epsilon^1$}\label{appendix_table_eps1}
\end{table}
\begin{table}[ht]
\def\arraystretch{2}
\begin{tabular}{c|c|c}
Term & Inner solution & Outer solution\\\hline \hline
$r^{-3/2}\sin\frac{3 \theta }{2}$ & $\frac{f^2}{8 }$ &$-\frac{q_2\; i e^{i\pi/4}}{2\sqrt{\pi } \alpha^{1/2}  \;(-\omega) ^{3/2}}$\\ \hline $r^{-1/2}\sin{\frac{\theta}{2}}$ & $\frac{1}{2} B^{(3/2)} f-A^{(3/2)}\frac{ 3f^2}{8}+\frac{ V f f_t}{4 \alpha ^2}$ & $-\frac{q_2\; i e^{i\pi/4}}{2\sqrt{\pi } \alpha^{1/2}  \;(-\omega) ^{3/2}} \left(\frac{3 \, i\, V \;\omega }{2 \alpha ^2}+\widehat{\Pi}\right)+\frac{q_1 e^{i\pi/4}}{\sqrt{-\pi\omega } \alpha^{1/2}}$\\ \hline
$r^{-1/2}\sin\frac{5 \theta }{2}$ & $-\frac{ V f f_t}{8 \alpha ^2}$ &$\frac{q_2\; i e^{i\pi/4}}{2\sqrt{\pi } \alpha^{1/2}  \;(-\omega)^{3/2}}\frac{i V \omega  }{2 \alpha ^2}$\\ \hline
$r^{1/2}\sin{\frac{\theta}{2}}$ 
	&\parbox{5cm}{\vspace{2mm} $\frac{15}{8} f^2 A^{(5/2)}-\frac{3}{2} f B^{(5/2)}+C^{(5/2)}\\ +\frac{ V f B_t^{(3/2)}}{2 \alpha ^2}-\frac{ 3A^{(3/2)} V f f_t}{4 \alpha ^2}+\frac{f f_{tt}}{4 \alpha ^4}-\frac{f f_{zz}}{4 \alpha ^2}$\vspace{2mm}} 
	& \parbox{5cm}{\vspace{2mm} $-\frac{q_2\; i e^{i\pi/4}}{2\sqrt{\pi } \alpha^{1/2}  \;(-\omega) ^{3/2}}\Big(\frac{k^2}{\alpha ^2}-\frac{\omega ^2}{\alpha ^4}-\frac{7 V^2 \omega ^2}{8 \alpha ^4}\\
		+\frac{5 i V \omega \widehat{\Pi}}{2 \alpha ^2}\Big)+\frac{q_1 e^{i\pi/4}}{\sqrt{-\pi\omega } \alpha^{1/2}}\left(\frac{i V \omega }{2 \alpha ^2}+\widehat{\Pi}\right)$\vspace{2mm}}\\ \hline
$r^{1/2}\sin\frac{3 \theta }{2}$ 
	&\parbox{5cm}{\vspace{2mm} $-B_t^{(3/2)}\frac{ V f}{4 \alpha ^2}-B^{(3/2)}\frac{ V f_t}{4 \alpha ^2}\\
		+A^{(3/2)}\frac{ 3 V ff_t}{8 \alpha ^2}	-\frac{\left(1+V^2\right) \left(f_t^2+f f_{tt}\right)}{8 \alpha ^4}\\
		+\frac{f f_{zz}}{8 \alpha ^2}+\frac{f_z^2}{8 \alpha ^2}$\vspace{2mm}} 
	&\parbox{5cm}{\vspace{2mm} $-\frac{q_2\; i e^{i\pi/4}}{2\sqrt{\pi } \alpha^{1/2}  \;(-\omega) ^{3/2}}\Big(-\frac{k^2}{2 \alpha ^2}+\frac{\omega ^2}{2 \alpha ^4}+\frac{3 V^2 \omega ^2}{4 \alpha ^4}\\
	-\frac{iV\omega \widehat{\Pi}}{2 \alpha ^2}\Big)-\frac{q_1 e^{i\pi/4}}{\sqrt{-\pi\omega } \alpha^{1/2}} \left[\frac{i V \omega  }{2 \alpha ^2}\right]$\vspace{2mm}}\\ \hline
$r^{1/2} \sin\frac{7 \theta }{2}$ & $\frac{V^2 \left(f_t^2+f f_{tt}\right)}{32 \alpha ^4}$& $-\frac{q_2\; i e^{i\pi/4}}{2\sqrt{\pi } \alpha^{1/2}  \;(-\omega) ^{3/2}} \left[
-\frac{V^2 \omega ^2 }{8 \alpha ^4}\right]$
\end{tabular}
\caption{Terms of order $\epsilon^2$}\label{appendix_table_eps2}
\end{table}

The zeroth order matching in Table~\ref{appendix_table_eps0} yields $A^{(3/2)} = \frac{m}{3}$ and $A^{(5/2)} = \frac{n}{5}$. The 1\textsuperscript{st}-order matching is given in Table \ref{appendix_table_eps1}. The first row yields an expression for $q_0$; $\frac{q_0 e^{i\pi/4}}{\sqrt{-\pi\omega } \alpha^{1/2}}=\frac{f }{2}$. The second row translates into 
\begin{equation}\label{appendix_eq_B32}
B^{(3/2)}=\frac{m}{2} f +\frac{V }{4\alpha^2}f_t+\frac{1}{2}\Pi[f]\,.
\end{equation}
 
Given the expressions for $A^{(3/2)},A^{(5/2)},q_0$ and $B^{(3/2)}$, the third, fourth and sixth rows of Table \ref{appendix_table_eps1} become identities. This is a consistency check that our calculations have not contained a mistake. The fifth row yields new information:
\begin{equation}\label{appendix_eq_B52}
B^{(5/2)}=\frac{n}{2}f-\frac{V^2}{16\alpha^4}f_{tt}-\frac{V}{4\alpha^2}\Pi[f_t]\,.
\end{equation}

The 2\textsuperscript{nd}-order matching in Table \ref{appendix_table_eps2} provides the three remaining coefficients $q_1,q_2$ and $C^{5/2}$. The first row of the table translates into $-\frac{q_2\; i e^{i\pi/4}}{2\sqrt{\pi } \alpha^{1/2}  \;(-\omega) ^{3/2}}=\frac{f^2}{8 }$. The determination of $q_2$ makes the third and fifth rows of Table \ref{appendix_table_eps2} an identity. The second row gives $q_1$ through the equation 
\begin{equation}
\frac{q_1 e^{i\pi/4}}{\sqrt{-\pi\omega } \alpha^{1/2}}=\frac{1}{2}(\frac{m}{2} f +\frac{V }{4\alpha^2}f_t+\frac{1}{2}\Pi[f]) f-\frac{ m f^2}{8}+\frac{ V f f_t}{4 \alpha ^2}  -\frac{1}{8}\left(\frac{3 V f f_t  }{ \alpha ^2}+\Pi[f^2]\right)\,,
\end{equation}
or
\begin{equation}
\frac{q_1 e^{i\pi/4}}{\sqrt{-\pi\omega } \alpha^{1/2}}=\frac{ m f^2}{8} +\frac{1}{4}\Pi[f] f-\frac{1}{8} \Pi[f^2]\,.
\end{equation}
The fourth row of Table \ref{appendix_table_eps2} is the center of this calculation since it would give the 2\textsuperscript{nd}-order correction for the energy-release-rate $G$. Substituting the expressions for $A^{(5/2)}$ and $B^{(5/2)}$ the inner solution cell of the fourth row is 
\begin{equation}
-\frac{3}{8} n f^2+ \frac{8+7 V^2}{32\alpha^4}ff_{tt}+\frac{5 V }{8\alpha^2}f\Pi[f_t]+C^{(5/2)} -\frac{f f_{zz}}{4 \alpha ^2}\,,
\end{equation}
while the outer solution cell is
\begin{equation}
\begin{split}
&-\frac{1}{4\alpha ^2}(f_z^2+f f_{zz})+\frac{(8+7 V^2)}{32\alpha ^4}(f_t^2+f f_{tt})+\frac{  V \Pi[f f_t]}{2 \alpha ^2}+\frac{ m V f f_t}{8\alpha ^2}\\
& +\frac{V}{8\alpha ^2}(f\Pi[f_t] +f_t\Pi[f] )
+\frac{ m \Pi[f^2]}{8} +\frac{1}{4}\Pi[\Pi[f] f]-\frac{1}{8} \Pi[\Pi[f^2]]\,.
\end{split}
\end{equation}
Since $\widehat{\Pi}^2 = -\frac{1}{\alpha^4}(\omega^2-\alpha^2 k^2)$, equating these two expressions yields a formula for the last remaining coefficient $C^{(5/2)}$:
\begin{equation}\label{appendix_eq_C52}
\begin{split}
&C^{(5/2)} = \frac{3}{8} n f^2+\frac{7 V^2}{32\alpha ^4}f_t^2-\frac{1}{4\alpha^4}f f_{tt} -\frac{V}{2\alpha^2}f\Pi[f_t]
+\frac{1}{4\alpha ^2}ff_{zz}+\frac{  V \Pi[f f_t]}{2 \alpha ^2}\\
&+\frac{ m V f f_t}{8\alpha ^2}
+\frac{V}{8\alpha ^2}f_t\Pi[f] 
+\frac{ m \Pi[f^2]}{8} +\frac{1}{4}\Pi[\Pi[f] f]\,.
\end{split}
\end{equation}

\subsection{Calculation of the energy-release-rate}

The previous sections derived the asymptotic field $ u$ up to the 2\textsuperscript{nd}-order in the perturbation to the crack front. The expressions derived in the last section for the undetermined coefficients of the inner solution can now be inserted into Eqs.~(\ref{appendix_eq_inner_eps0_polar}-\ref{appendix_eq_inner_eps2_polar}) to find the explicit dependence of $ u$ on $f$. To calculate the energy-release-rate $G$, it is simpler to use Eqs.~(\ref{appendix_eq_Phi12}-\ref{appendix_eq_Phi52}). Then, the displacement and stress intensity factors are given respectively by
\begin{align}
K^u&=\lim\limits_{X\uparrow f(z,t)}\epsilon ^{-1/2}\left(X-f(z,t)\right)^{-1/2} \sqrt{\frac{\pi}{2}}\mu U (X,0,z,t)\,,\\
 K^\sigma &=\sqrt{2\pi}\lim\limits_{X\downarrow f(z,t)}\epsilon ^{-1/2}\sqrt{2\pi\left(X-f(z,t)\right)} \mu U_Y (X,0,z,t)\,.
\end{align} 

Some algebra leads to the two 2\textsuperscript{nd}-order expressions 
\begin{align}
K^u=& K^u_0 \left(1+\epsilon  \left(B^{(3/2)}+\frac{V f_t}{4 \alpha ^2}\right)\right.\nonumber\\
&\left.+\epsilon ^2 \left(C^{(5/2)}+\frac{V B^{(3/2)} f_t}{4 \alpha ^2}+\frac{f_t^2}{8 \alpha ^4}+\frac{3 V^2 f_t^2}{32 \alpha ^4}-\frac{f_z^2}{8 \alpha ^2}\right)\right)\,,\\
K^\sigma=&\sqrt{\frac{\pi }{2}} \alpha  K^u_0 \left(1  +\epsilon  \left( B^{(3/2)}-\frac{3 V f_t}{4 \alpha^2 }\right)\right.\nonumber\\
&\left.+\epsilon ^2 \left( C^{(5/2)}-\frac{3  V B^{(3/2)} f_t}{4 \alpha^2 }-\frac{3  f_t^2}{8 \alpha ^4}-\frac{5 V^2 f_t^2}{32 \alpha^4}+\frac{3  f_z^2}{8 \alpha^2 }\right)\right)\,.
\end{align}
The energy-release-rate is given by their product
\begin{equation}
\begin{split}
G \propto K^u K^\sigma =\alpha   (K^u_0)^2 \left(1+\epsilon  \left(2 B^{(3/2)}-\frac{V f_t}{2 \alpha ^2}\right)\right.\\
\left.+\epsilon ^2 \left((B^{(3/2)})^2+2  C^{(5/2)}-\frac{V B^{(3/2)} f_t}{\alpha ^2}-\frac{(1+V^2)f_t^2}{4 \alpha ^4}+\frac{f_z^2}{4 \alpha ^2}\right)\right)\,.
\end{split}
\end{equation}
Using Eqs.~(\ref{appendix_eq_B32},\ref{appendix_eq_C52}) derived in the last section for $B^{(3/2)}$ and $C^{(5/2)}$, we write 
\begin{equation}\label{appendix_eq_G_2nd}
G = G_r g(V)(1+\epsilon \delta G^{(1)}+\epsilon^2 \delta G^{(2)} +\order{f^3})\,,
\end{equation}
where the 1\textsuperscript{st}-order correction is
\begin{equation}\label{appendix_eq_g1}
\delta G^{(1)} = mf+\Pi[f]
\end{equation} 
and the 2\textsuperscript{nd}-order correction is
\begin{equation}\label{appendix_eq_g2}
\begin{split}
\delta G^{(2)}=&\frac{1}{4} m^2 f^2+\frac{3}{4} n f^2+\frac{1}{2} m f \Pi [f]+\frac{1}{4} m \Pi \left[f^2\right]\\
&+\frac{1}{4} \Pi [f]^2 +\frac{1}{2} \Pi [f \Pi [f]]-\frac{V f \Pi \left[f_t\right]}{\alpha ^2}+\frac{V \Pi \left[f f_t\right]}{\alpha ^2}-\frac{f_t^2}{4 \alpha ^4}-\frac{f f_{tt}}{2 \alpha ^4}+\frac{f_z^2}{4 \alpha ^2}+\frac{f f_{zz}}{2 \alpha ^2}\,
\end{split}
\end{equation}

The 1\textsuperscript{st}-order correction~(\ref{appendix_eq_g1}) was derived by \cite{norrisMultiplescalesApproachCrackfront2007}, where it was shown that the term $m f$ results in wave dispersion. This result is possibly related to the observations of small amplitude wave dispersion reported in \cite{Fineberg.03}. Re-introducing dimensions, the coefficient $m$ scales as $1/l$, and therefore the term $mf\sim\order{f/l}$ becomes negligible when $f\ll l$. The term $\Pi[f]$ remains as it does not depend on sample geometry or loading conditions. In the following, we will assume  $f/l\rightarrow 0$ and put $m=n=0$. The expansion of the energy-release-rate~(\ref{appendix_eq_G_2nd}) can be rewritten as a product of dynamical and ``historical'' contributions, as in Eq.~(\ref{appenidx_eq_G_decomposed}). To show that we follow~\cite{Morrissey.00} and convert $\widehat{\Pi}(k,\omega)$ to the time domain using the identity
\begin{equation}
\int \!\mathrm{d}t\, e^{-i\omega t} \left(\partial_t +\frac{p \Theta(t)J_1(pt)}{t}\right)=\sqrt{p^2 - \omega^2}\,,
\end{equation}
where $\Theta$ is the Heaviside function and $J_1$ is the 1\textsuperscript{st}-order Bessel function. Then, $\Pi(k,t)  = -\alpha^{-2}(\partial_t + \Psi(k,t))$, where $\Psi(k,t) = \alpha k (\Theta(t)J_1(\alpha k t)/t)*$, with $*$ representing a convolution in time. The appearance of the Heaviside function in $\Psi[f]$ signifies the dependence of real space functional $\Pi[f](z,t)$ on the history of the front. With this transformation Eq.~(\ref{appendix_eq_G_2nd}) becomes (setting $m=n=0$ and $\epsilon=1$)
\begin{align}\label{appendix_eq_G_2nd_time_domain}
G = & G_r g(V)\left\{1-\frac{f_t}{\alpha^{2}}-\frac{1}{\alpha^{2}}\Psi[f]+\frac{1}{4\alpha^4}\Psi[f]^2 +\frac{1}{2\alpha^4}\Psi[f\Psi[f]]+\right.\nonumber\\
&\left.\frac{1-2 V}{2\alpha^4}\Psi[ff_t]+\frac{1+2 V}{2\alpha^4}f\Psi[f_t]+\frac{1}{\alpha^4}f_t\Psi[f]+\frac{1-2 V}{2\alpha^4}f_t^2+\frac{1}{4\alpha^2}f_z^2+\frac{1}{2\alpha^2}ff_{zz}\right\}\,.
\end{align}
To write this expression under the form given by Eq.~(\ref{appenidx_eq_G_decomposed}), we develop $V_\perp = V+f_t-\frac{V}{2}f_z^2+\order{f^3}$ so that
\begin{equation}
g(V_\perp) = g(V)\left(1-\frac{f_t}{\alpha^2}+\frac{V}{2\alpha^2}f_z^2+\frac{1-2 V}{2\alpha^4}f_t^2 +\order{f^3}\right)\,.
\end{equation}
Inserting this expression into Eq.~(\ref{appenidx_eq_G_decomposed}) we have
\begin{equation}\label{appendix_eq_G_decomp_developed}
G = G_r g(V)\left(1-\frac{f_t}{\alpha^2}+H^{(1)}[f]+\frac{V}{2\alpha^2}f_z^2+\frac{1-2 V}{2\alpha^4}f_t^2-\frac{f_t}{\alpha^2}H^{(1)}[f]+H^{(2)}[f]\right)\,.
\end{equation}
A direct comparison between Eqs.~(\ref{appendix_eq_G_2nd_time_domain}) and~(\ref{appendix_eq_G_decomp_developed}) then yields  Eq.~(\ref{appendix_eq_h1}) for $H^{(1)}[f]$ and
\begin{equation}
H^{(2)}[f] =  \frac{1}{4\alpha^4}\Psi[f]^2 +\frac{1}{2\alpha^4}\Psi[f\Psi[f]]+\frac{1-2 V}{2\alpha^4}\Psi[ff_t]+\frac{1+2 V}{2\alpha^4}f\Psi[f_t]+\frac{1-2 V}{4\alpha^2}f_z^2+\frac{1}{2\alpha^2}ff_{zz}\,.
\end{equation}
The expression for $H^{(2)}[f]$ can be further simplified through integration by parts in the time domain. Using
\begin{equation}
\begin{split}
\Psi[f_t] &= \alpha k \int_{-\infty}^t \frac{J_1(\alpha k (t-t'))}{t-t'}f_t(k,t')dt' \\
&= \frac{\alpha^2 k^2}{2}f(k,t)-\alpha^2 k^2\int_{-\infty}^t \frac{J_2(\alpha k (t-t'))}{t-t'}f(k,t')dt'\,,
\end{split}
\end{equation}
the 2\textsuperscript{nd}-order correction $H^{(2)}[f]$ is then given by Eq.~(\ref{appendix_eq_h2}).

\section{Nonlinear perturbation of Mode I elastic crack fronts}\label{vectorCracks}

In this section, we consider dynamic crack fronts in materials described by 3D elastodynamics which involves three displacement components $u_i(x,y,z,t)$ and a system of three scalar wave equations coupled by boundary conditions. The wave equations are  a consequence of Newton's 2\textsuperscript{nd} law
\begin{equation}
\frac{\partial \sigma_{ix}}{\partial x}+\frac{\partial \sigma_{iy}}{\partial y}+\frac{\partial \sigma_{iz}}{\partial z}=\rho \frac{\partial^2 u_i}{\partial t^2}\,.\label{eq:bulk}
\end{equation}
The 3D stress $\sigma_{ij}$  is linearly related to the strain tensor through the Young modulus $E$ and the Poisson ratio $\nu$. Under pure Mode~I loading, the boundary conditions comprise tensile loads applied at the remote boundaries and traction-free conditions on the crack faces, i.e.
\begin{equation}
\sigma_{ij}n_j=0\,.
\end{equation}

The perturbation scheme of planar crack fronts follows the PhD thesis of \cite{RamanathanThesis}. Parts of the calculation, that appear in the thesis, are reproduced below for completeness. The explicit 2\textsuperscript{nd}-order expressions and the time-dependent formulation are the main contributions of this section. To compute the expansion of $G$ in $f$, Ramanathan's calculation aims at obtaining an asymptotic solution for the elastic fields of a running in-plane crack with a crack front defined by $h(z,t) = Vt + f(z,t)$ (Fig. 1). The solution is to be obtained as a perturbation series around a straight front $f(z,t)=0$. More specifically, the $x$-origin of the system of coordinates is locally translated to the instantaneous position of the crack front $(X\equiv x-h(z,t),y,z)$. Then the contribution of $f$ in the elastic field components is ``eliminated". This latter condition is fulfilled by introducing $f(z,t)$ at the functional level through the transformation
\begin{equation}\label{fieldsdisplaced0}
\begin{split}
&u_i(X,y,z,t) = e^{  f(z,t)\partial_X}U_i(X,y,z,t)\,,\\
&\sigma_{ij}(X,y,z,t) = e^{  f(z,t)\partial_X}\Sigma_{ij}(X,y,z,t)\,.
\end{split}
\end{equation}
The advantage of this transformation is that the new fields $U_i$ and $\Sigma_{ij}$ still follow the same constitutive linear elastic relations as $u_i$ and $\sigma_{ij}$. Moreover, Eq.~(\ref{eq:bulk}) becomes
\begin{equation}
\frac{\partial \Sigma_{ix}}{\partial X}+\frac{\partial \Sigma_{iy}}{\partial y}+\frac{\partial \Sigma_{iz}}{\partial z}=\rho \left(V^2\frac{\partial^2 U_i}{\partial X^2}-2V\frac{\partial^2 U_i}{\partial X\partial t}+\frac{\partial^2 U_i}{\partial t^2}\right)\,.
\end{equation}
Using the symmetry of Mode~I loading, the boundary conditions for the fields $U_i$ and $\Sigma_{ij}$ on the plane $y=0$ are given by
\begin{eqnarray}
\Sigma_{xy}(X,0,z,t) &=& \Sigma_{yz}(X,0,z,t)=0\,,\label{eq:bcSigma}\\
U_y(X>0,0,z,t) &=& 0\,,\; \Sigma_{yy}(X<0,0,z,t)=0\label{eq:bcU}\,.
\end{eqnarray}
The fields $\Sigma_{ij}$ and $U_i$ should be computed from the solution of the elastodynamic equations in the reference frame $(X,y,z)$ with the prescribed boundary conditions. This is done by using the general relations obtained by~\cite{geubelleSpectralMethodThreedimensional1995} between the Fourier components of stress and displacement fields in the fracture plane. Specifically, the stress component $\Sigma_n(X,z,t)\equiv \Sigma_{yy}(X,y=0,z,t)$ and the displacement component $U_n(X,z,t)=U_y(X,y=0^+,z,t)$ are related through
\begin{equation}
\widehat{\Sigma}_{n}(q,k,\omega) = \widehat{\Upsilon}_{yy}(q,k,\Omega)\widehat{U}_n(q,k,\omega)\,,
\end{equation}
where the Fourier transforms are defined such that $\widehat{F}(q,k,\omega)  = \int\,dX\, dz\, dt\, e^{-iqX-ikz-i\omega t}F(X,z,t)$ and 
\begin{equation}\label{eq:upsilon}
\hat{\Upsilon}_{yy}(q,k,\Omega) = -\mu  \frac{ p\,R\left(\frac{\Omega}{p}\right)}{\frac{\Omega^2}{b^2 p^2}\sqrt{1-\frac{\Omega^2}{a^2 p^2}}}\,,
\end{equation}
where $\mu$ is the shear modulus, $\Omega = \omega- q V$, $p=\sqrt{q^2+k^2}$ and $R(\zeta)$ is the Rayleigh function given by
\begin{equation}\label{eq:frayleigh}
    R(\zeta)
 = 4\sqrt{1-\frac{\zeta^2}{a^2 }}\sqrt{1-\frac{\zeta^2}{b^2 }}-\left(2-\frac{\zeta^2}{b^2 }\right)^2\,.
 \end{equation}
Here $a$ and $b$ are the longitudinal and shear wave speeds.

It is left to explicitly compute $\Sigma_{yy}$ and $U_y$ by satisfying the boundary condition~(\ref{eq:bcU}). Notice that $\Sigma_{n}(X<0,z,t)=0$ (resp. $U_{n}(X<0,z,t)=0$) is equivalent to $\widehat\Sigma_{n}(q,k,\omega)$ (resp; $\widehat U_{n}(q,k,\omega)$) having its analytical domain encompassing the upper (resp. lower) $q$-half-plane. Then, given a decomposition $\widehat{\Upsilon}_{yy}=\widehat{\Upsilon}^+\widehat{\Upsilon}^-$ where $\widehat{\Upsilon}^+$ is analytical for $\mbox{Im}(q)>0$ and $\widehat{\Upsilon}^-$ is analytical for $\mbox{Im}(q)<0$, there exists an analytical function $\widehat{W}(q,k,\omega)$ for all $q$'s such that $\widehat{\Sigma}_{n} = \widehat{W}\widehat{\Upsilon}^+$ and $\widehat{U}_n = \widehat{W}/\widehat{\Upsilon}^-$. Finally, the crux of the method lies in demanding that the fields have the correct physical singularity locally at the front such that
\begin{equation}\label{physicaldemand}
\begin{split}
&\sigma_{yy}(X,y=0,z,t) \rightarrow K^\sigma(z,t) ( 2 \pi X)^{-1/2}\;\; \text{ as }\;\; X\rightarrow 0^+ \,,\\&
u_y(X,y=0^+,z,t) \rightarrow \frac{1-\nu^2}{E}K^u(z,t) (-2 X/\pi)^{1/2}\;\; \text{ as }\;\; X\rightarrow 0^-\,,
\end{split}
\end{equation}
where $K^\sigma$ and $K^u$ are the stress and displacement intensity factors respectively.
For a straight, unperturbed, crack front, $K^\sigma = K^\sigma_0 = K^\sigma_r k(V)$ and $K^u = K^u_0 =2 A_I(V)K^\sigma_0$ where the rest SIF $K^\sigma_r$ is determined by the loading conditions~\citep{Freund.90}  (see also section \ref{subsection_Mode_I_summary} for definitions of $k(V)$ and $A_I(V)$). For a curved crack front, the demand, Eq.~(\ref{physicaldemand}) will be met by eliminating all the higher order singularities that appear on the right-hand side of Eq.~(\ref{fieldsdisplaced0}). In $q$-space, this demand imposes that the field $\widehat{u}_y$ shall not contain any power of $q$ higher than $q^{-3/2}$. 

The energy-release-rate can be directly expressed using $K^{\sigma,u}$. Following \cite{Freund.90}, we construct a small parallelopiped $P = (\delta x,\delta y,\delta z)$ around the point on the crack front $x=h(z,t)$ that has two sides of parallel to the $x-z$ plane, two sides parallel to the local front tangent and $\hat{y}$, and two sides parallel to the $x-y$ plane. The parallelopiped also travels at velocity $\dot{h}=V+f_t$ along the $x$-axis. With these definitions, the local energy-release-rate is given by 
\begin{equation}
    G(z,t) = \lim_{P\rightarrow 0}\left\{\frac{1}{\delta z}\frac{1}{\dot{h}}\int_{P}\!\mathrm{d}S\, \sigma_{ij}\hat{n}_j\dot{u}_i \right\}\,
\end{equation}
where $\hat{n}_i$ is the local normal to $P$ and $\mathrm{d}S$ is a surface element. Since only $\sigma_{yy}$ and $\dot{u}_y$ have singular $X^{-1/2}$ behavior close to the crack front, the above integral simplifies into
\begin{equation}
    G(z,t) = \lim_{\delta x\rightarrow 0}\lim_{\delta y\rightarrow 0}\left\{\frac{2}{\dot{h}}\int_{h(z,t)-\delta x}^{h(z,t)+\delta x}\!\mathrm{d}x\, \sigma_{yy}\dot{u}_y \right\} = \lim_{\delta x\rightarrow 0}\lim_{\delta y\rightarrow 0}\left\{2\int_{-\delta x}^{\delta x}\!\mathrm{d}X\, \Sigma_{yy}\pd{U_y}{X} \right\}\,
\end{equation}
which leads to
\begin{equation}\label{eq:G_equal_KuKs}
    G(z,t) = \frac{1-\nu^2}{2E}K^u K^\sigma\,.
\end{equation}
Alternative formulations of the 3D path-independent energy flux integral give the same result~\citep{amestoyDefinitionLocalPath1981,doddsNumericalEvaluation3D1988,erikssonDomainIndependentIntegral2002,leguillonAttemptExtend2D2014,elkabirNewAnalyticalGeneralization2018}. 
\subsection{Implementation}

To compute the energy-release-rate, we apply the Fourier transform $F(q) = \int\!\mathrm{d}X\,e^{iqX} F(X)$ to Eqs.~(\ref{fieldsdisplaced0}) ,
\begin{equation}\label{fieldsdisplaced}
\begin{split}
&\widetilde{u}_y(q,y=0^+,z,t) = e^{i qf(z,t)}\widetilde{U}_n(q,z,t)\,,\\
&\widetilde{\sigma}_{yy}(q,y=0,z,t) = e^{i qf(z,t)}\widetilde{\Sigma}_{n}(q,z,t)\,.
\end{split}
\end{equation}
Consider the displacement first. The zeroth order solution in the vicinity of the straight crack front is given by  
\begin{equation}
{u}_0(X,y=0^+,z,t) = \frac{1-\nu^2}{E}K_0^u\sqrt{\frac{-2X}{\pi}}\Theta(-X)\,,
\end{equation}
which transforms into 
\begin{equation}
\widetilde{u}_0(q,y=0^+,z,t) =\kappa^u q^{-3/2}\,,
\end{equation}
in the sense that $q$ has a vanishing positive imaginary part, and $\kappa^u = i^{3/2}(1-\nu^2)(\sqrt{2}E)^{-1}K_0^u$. The next orders will be given by $\widehat{W}/\widehat{\Upsilon}^-$, such that
\begin{equation}\label{UWY}
\widetilde{u}_y(q,y=0^+,z,t)=e^{i qf(z,t)}\left[\widetilde{u}_0(q,y=0^+,z,t)+\int \frac{dk}{2\pi}\frac{d\omega}{2\pi}e^{ikz+i\omega t} \frac{\widehat{W} (q,k,\omega)}{\widehat{\Upsilon}^-(q,k,\Omega)}\right]\,,
\end{equation}   
where $\widehat{W}(q,k,\omega)$ is an analytical function of $q$. Now we expand  all functions in $q$~:
\begin{equation}
\begin{split}
&e^{i q f(z,t)} = \sum_{j=0}^{\infty} \frac{(i q)^j}{j!}f^j(z,t)\,,\\
& \widetilde{W}(q,z,t) =\sum_{r=1}^{\infty} (i q)^{r-1}W_r(z,t)\,,\\
&\frac{1}{\widetilde{\Upsilon}^-(q,z,t)} = \kappa^u q^{-1/2} \left(1+\sum_{n=1}^{\infty}\frac{\Lambda_n^-(z,t)}{(iq)^n}\right)\,.  
\end{split}
\end{equation}
Multiplying~(\ref{UWY}) by $q^{3/2}/\kappa^u$ we obtain
\begin{equation}\label{Uexpansion0}
\frac{q^{3/2}}{\kappa^u}\widetilde{u}_y (q,y=0^+,z,t) =  \sum_{j=0}^\infty(iq)^j\frac{f^j}{j!}\left[ 1-i\sum_{r=1}^\infty(iq)^{r} W_r-i\sum_{r=1}^\infty\sum_{n=1}^\infty(iq)^{r-n}\Lambda_n^-\star W_r\ \right]\,,
\end{equation}
where $\star$ denotes convolution in $z$ and $t$. Eq.~(\ref{Uexpansion0}) can be simplified into
\begin{equation}\label{Uexpansion}
\begin{split}
&\frac{q^{3/2}}{\kappa^u}\widetilde{u}_y (q,y=0^+,z,t) = 1 - i\sum_{n=1}^\infty\sum_{r=1}^n\frac{f^{(n-r)}}{(n-r)!} (\Lambda_n^-\star W_r)\\
& +\sum_{m=1}^\infty(iq)^m\left[\frac{f^m}{m!} - i\sum_{r=1}^{m} \frac{f^{(m-r)}}{(m-r)!} W_r - i\sum_{n=1}^\infty\sum_{r=1}^{m+n} \frac{f^{(m+n-r)}}{(m+n-r)!} (\Lambda_n^-\star W_r)\right]\,.
\end{split}
\end{equation}
To meet the physical demand given by Eq.~(\ref{physicaldemand}), we need to equate in Eq.~(\ref{Uexpansion}) the coefficients of $(iq)^m$ with $m\geq1$ to zero. Therefore we get
\begin{equation}\label{Wrecursive}
i\sum_{n=1}^{m} \frac{f^{(m-n)}}{(m-n)!}W_n+i\sum_{n=1}^\infty\sum_{r=1}^{m+n}\frac{f^{(m+n-r)}}{(m+n-r)!}(\Lambda_n^-\star W_r)=\frac{f^m}{m!}\,.
\end{equation}
Eq.~(\ref{Wrecursive}) should be solved for $W_m$ order by order in $f$. Notice that Eq.~(\ref{Wrecursive}) shows that each $W_r$ has at least one term of the order $f^r$. Therefore, $W_r$ can be written as
\begin{equation}
W_r(z,t) = -i\sum_{j=r}^\infty w_{r,j}(z,t)\,,
\end{equation}
where $w_{r,j}$ is of the order $O(f^j)$. Plugging this expansion in Eq.~(\ref{Wrecursive}) we find
\begin{equation}
\begin{split}
&\sum_{n=1}^{m} \frac{f^{(m-n)}}{(m-n)!} w_{n,n}= \frac{f^m}{m!}\,\quad m\geq 1\,,\\
&\sum_{n=1}^{m} \frac{f^{(m-n)}}{(m-n)!} w_{n,l+n-m}+\sum_{n=1}^{l-m}\sum_{r=1}^{m+n}\frac{f^{(m+n-r)}}{(m+n-r)!}(\Lambda_n^-\star w_{r,l+r-m-n})= 0\,\quad l> m\geq 1\,.
\end{split}
\end{equation}
It is easy to show that this linear system of equations can be solved recursively order by order. For the orders of interest, we find
\begin{equation}
\begin{split}
    &w_{1,1} = f\,,\\
    &w_{2,2} = \frac{f^2}{2} - f w_{1,1} = -\frac{f^2}{2}\,,\\
    &w_{1,2} = -f(\Lambda_1^-\star w_{1,1}) -\Lambda_1^-\star w_{2,2} = -f(\Lambda_1^-\star f) +\Lambda_1^-\star \frac{f^2}{2}\,.
\end{split}
\end{equation}

It is left to express the SIF as a series in $f$. The expansion of the SIF comprises the remaining non-zero terms in Eq.~(\ref{Uexpansion}). Expressing these terms with $w_{r,j}$,
\begin{equation}
\frac{K^u}{K^u_0} = 1-\sum_{p=1}^\infty{\cal C}_p =  1-\sum_{p=1}^\infty\sum_{n=1}^p\sum_{r=1}^n\frac{f^{(n-r)}}{(n-r)!}  (\Lambda_n^-\star w_{r,p+r-n})\,.
\end{equation}
For the orders of interest, we find
\begin{equation}
\begin{split}
&{\cal C}_1 = \Lambda_1^-\star f\,,\\
&{\cal C}_2 =  -\Lambda_1^- \star\{f(\Lambda_1^-\star f)\}+\Lambda_1^-\star\Lambda_1^-\star\frac{f^2}{2} +f\{\Lambda_2^-\star f\}-\Lambda_2^-\star \frac{f^2}{2}\,.
\end{split}
\end{equation}
Finally, the expression for the displacement intensity factor is
\begin{equation} \label{eq:Ku_expansion}
\frac{K^u}{K^u_0} = 1-\Lambda_1^-\star f +\Lambda^-_1\star\{f(\Lambda_1^-\star f)\}-\Lambda^-_1\star\Lambda_1^-\star \frac{f^2}{2} - f(\Lambda_2^-\star f)+\Lambda_2^-\star\frac{f^2}{2}\,.
\end{equation}

A similar calculation shows that the SIF is
\begin{equation} \label{eq:Ks_expansion}
\frac{K^\sigma}{K^\sigma_0} = 1-\Lambda_1^+\star f +\Lambda^+_1\star\{f(\Lambda_1^+\star f)\}-\Lambda^+_1\star\Lambda_1^+\star \frac{f^2}{2} - f(\Lambda_2^+\star f)+\Lambda_2^+\star\frac{f^2}{2}\,.
\end{equation}
where the functions $\Lambda_n^+$ are given by the expansion
\begin{equation}
\widetilde{\Upsilon}^+(q,z,t) = (2i)^{-1/2} K_0^\sigma q^{1/2} \left(1+\sum_{n=1}^{\infty}\frac{\Lambda_n^+(z,t)}{(iq)^n}\right)\,.
\end{equation}
To obtain explicit expressions for the SIF and the energy-release-rate, one should perform the decomposition of $\widehat{\Upsilon}_{yy}(q,k,\omega)$ and its expansion in powers of $q$. This is done in the next section.

\subsection{Decomposition of \texorpdfstring{$\widehat{\Upsilon}_{yy}(q,k,\omega)$}{Upsilonyy}}\label{Dec_Ups}

According to our plan, we seek a Wiener-Hopf decomposition of $\widehat{\Upsilon}_{yy}(q,k,\omega)$ (Eq.~(\ref{eq:upsilon})) into a product of two functions $\widehat{\Upsilon}^+(q,k,\omega)$ and $\widehat{\Upsilon}^-(q,k,\omega)$ that are analytical for $\mbox{Im}\,q<0$ and $\mbox{Im}\,q>0$ respectively. Similarly to section 2.5 in \cite{Freund.90}, a function $T(\Omega/p)\propto \widehat{\Upsilon}_{yy} $ is sought that satisfies $T\rightarrow 1$ when $|q|\rightarrow \infty$ and $\log T = \log T^+ +\log T^-$, where $T^+$ ($T^-$) is analytical and non-zero in the lower (upper) half of the $q$-plane. To find $T$, let us analyze the Rayleigh function given by Eq.~(\ref{eq:frayleigh}). It is known that $R(\zeta)$ has a double root at $\zeta = 0$ and two simple roots at $\zeta = \pm c$, where $c<b<a$ is the Rayleigh velocity. To extract these roots from $\widehat{\Upsilon}_{yy}$, we define 
\begin{equation}\label{T}
T (\zeta)= \frac{V^2}{R(V)\gamma_c^2}\frac{R(\zeta)}{\zeta^2\alpha_c(\zeta)^2}\,,
\end{equation}
with $\alpha_c(\zeta) = \sqrt{1-\frac{\zeta^2}{c^2}}$ and $\gamma_c = 1/\sqrt{1-V^2/c^2}$. Since $T(V)=1$, we have  $T(\Omega/p)\rightarrow 1$ as $\abs{q}\rightarrow\infty$. Then, Eq.~(\ref{eq:upsilon}) is rewritten as
\begin{equation}
\widehat{\Upsilon}_{yy}(q,k,\omega) =  - \mu \frac{b^2 R(V)\gamma_c^2}{V^2} p \frac{\alpha_c(\Omega/p)^2}{\alpha_a(\Omega/p)}T(\Omega/p)\,,
\end{equation}
with $\alpha_a(\zeta) = \sqrt{1-\frac{\zeta^2}{a^2}}$.  Moreover, we notice that
\begin{equation}\label{eq:numden}
p \frac{\alpha_c(\Omega/p)^2}{\alpha_a(\Omega/p)} = \frac{q^2+k^2-(\omega-qV)^2/c^2}{\sqrt{q^2+k^2-(\omega-qV)^2/a^2}}\,.
\end{equation}
Both the denominator and the numerator in Eq.~(\ref{eq:numden}) contain similar quadratic expressions in $q$ and they may be decomposed in the same manner. For example, one has
\begin{equation}
q^2+k^2-\frac{(\omega-qV)^2}{c^2} = \frac{1}{\gamma_c^2}\left(q+\frac{V\omega\gamma_c^2}{c^2}-iQ_c\right)\left(q+\frac{V\omega\gamma_c^2}{c^2}+iQ_c\right)\,,
\end{equation}
where $Q_c = \gamma_c\sqrt{k^2-\frac{\omega^2\gamma_c^2}{c^2}}$. Finally, one has
\begin{equation}
\widehat{\Upsilon}_{yy}(q,k,\omega) = -\mu \frac{b^2 R(V)}{V^2} \frac{\left(q+\frac{V\omega\gamma_c^2}{c^2}-iQ_c\right)\left(q+\frac{V\omega\gamma_c^2}{c^2}+iQ_c\right)}{\frac{1}{\gamma_a}\sqrt{q+\frac{V\omega\gamma_a^2}{a^2}-iQ_a}\sqrt{q+\frac{V\omega\gamma_a^2}{a^2}+iQ_a}}T(\Omega/p)\,.
\end{equation}
and
\begin{eqnarray}\label{Upsilons}
\widehat{\Upsilon}^+ &=& \frac{1}{(2i)^{1/2}} K_0^\sigma \frac{q+\frac{V\omega\gamma_c^2}{c^2}-iQ_c}{\sqrt{q+\frac{V\omega\gamma_a^2}{a^2}-iQ_a}}T^+ \\ 
\widehat{\Upsilon}^- &=& \frac{2^{1/2} E}{i^{3/2}(1-\nu^2) K_0^u}\frac{q+\frac{V\omega\gamma_c^2}{c^2}+iQ_c}{\sqrt{q+\frac{V\omega\gamma_a^2}{a^2}+iQ_a}}T^-\,.
\end{eqnarray}

The decomposition of $\hat\Upsilon_{yy} $ reduces to resolving $T = T^+ T^-$. Let us identify the 
singularities of $T$ in the $q$ plane. $R(\zeta)$ has two branch cuts: namely 
$b<\zeta<a$ and $-a<\zeta<-b$. Since $\zeta = \frac{\omega -q V}{\sqrt{q^2+k^2}}$, in the $q$ plane these branch cuts are 
transformed into into linear segments from $q_a$ to $q_b$ in the upper $q$
plane and from $q_a^*$ to $q_b^*$ in the 
lower $q$ plane, where $q_a =- \frac{\omega V \gamma_a^2}{a^2}+iQ_a$,  $q_b = -\frac{\omega V \gamma_b^2}{b^2}+iQ_b$ 
and $Q_a = \gamma_a\sqrt{k^2-\frac{\omega^2\gamma_a^2}{a^2}}$,
$Q_b = \gamma_b\sqrt{k^2-\frac{\omega^2\gamma_b^2}{b^2}}$.

After \cite{RamanathanThesis}, we name the contour circling $[q_a,q_b]$ in the clockwise direction the  branch cut $C_+$ and
the contour circling $[q_a^*,q_b^*]$ in the clockwise direction the  branch cut $C_-$.
Then, $\log T $ satisfies the conditions for 
the Wiener-Hopf decomposition: it decays
to $0$ as $\abs{q}\rightarrow\infty$
everywhere and it has two finite branch cuts. Therefore, one has
\begin{eqnarray}
T^+ &=& \exp\left(\frac{1}{2\pi i}\oint_{C_+}\! \frac{\log T}{\xi -q}\,\mathrm{d}\xi\right) \\
T^- &=& \exp\left(\frac{1}{2\pi i}\oint_{C_-}\! \frac{\log T}{\xi -q}\,\mathrm{d}\xi\right)\,.
\end{eqnarray} 
Let's tackle the first integral. 
Writing it more explicitly we have
\begin{equation}
\frac{1}{2\pi i}\oint_{C_+}\! \frac{\log T}{\xi -q}\,\mathrm{d}\xi = \frac{1}{2\pi i}\oint_{C_+}\! \frac{\mathrm{d}\xi}{\xi -q}\,\log\left(R(\zeta)/B(\zeta)\right)\,,
\end{equation}
where $\zeta =\frac{\omega -\xi V}{\sqrt{\xi^2+k^2}}$ and
$B(\zeta) = V^{-2}R(V)\gamma_c^2\zeta^2(1-\zeta^2/c^2)$. Then, since $B(\zeta)$ has no 
poles or zeros inside the contours $C_\pm$, the
integral over its logarithm is trivially zero and we have
\begin{equation}
\log T^\pm = \frac{1}{2\pi i}\oint_{C_\pm}\! \frac{\log T}{\xi -q}\,\mathrm{d}\xi = \frac{1}{2\pi i}\oint_{C_\pm}\! \frac{\mathrm{d}\xi}{\xi -q}\,\log\left(R(\zeta)\right)\,.
\end{equation} 
Changing the integration variable to $\zeta = \zeta(\xi)$ such that
\begin{equation}
q_\pm (\zeta) = -\frac{\omega V}{\zeta^2-V^2}\pm i\abs{k}\frac{\sqrt{\zeta^2}\sqrt{\zeta^2-H}}{\zeta^2-V^2}\,,
\end{equation} 
where $H = V^2+\omega^2/k^2$. Here, the $+$ sign is taken for the $C_+$ integral 
and the $-$ sign is taken for the $C_-$ integral.
Then
\begin{equation}
\frac{\mathrm{d}\xi}{\xi-q} = \frac{q_\pm '(\zeta)\,\mathrm{d}\zeta}{q_\pm(\zeta)-q}\,.
\end{equation}
This change of variables copies the contour $C_+$ in the $q$-plane to a clockwise contour around the real interval $b<\zeta<a$ and the contour $C_-$ to a clockwise contour around $-a<\zeta<-b$. 
To evaluate the contour integrals, a choice for the branch cuts of $R(\zeta)$ has to be made. To ensure that the elastic fields maintain physical behavior far from the crack, $\alpha_b$ must always have a non-negative real part~\citep{geubelleSpectralMethodThreedimensional1995,RamanathanThesis}. For $\zeta=x+i\epsilon$ where $x>b$ and $|\epsilon|$ is arbitrarily small, we have $\alpha_b\simeq \pm i \sqrt{x^2/b^2-1}(1+i\epsilon x/(x^2-b^2))$. Hence, when $\epsilon>0$, we must choose the minus sign, and vice versa when $\epsilon<0$. Accordingly, the integral which follows the clockwise contour $C_+$ is
\begin{eqnarray}
\log T^+ = \frac{1}{2\pi i}\oint_{C_+}\! \frac{q_+ '(\zeta)\,\mathrm{d}\zeta}{q_+(\zeta)-q}\,\log\left(R(\zeta)\right) =  \nonumber \\
\frac{1}{2\pi i}\int_{b}^{a}\! \frac{q_+ '(\zeta)\,\mathrm{d}\zeta}{q_+(\zeta)-q}\log\left[-4i\sqrt{1-\frac{\zeta^2}{a^2}}\sqrt{\frac{\zeta^2}{b^2}-1}-(2-\frac{\zeta^2}{b^2})^2\right] + \nonumber\\
\frac{1}{2\pi i}\int_{a}^{b}\! \frac{q_+ '(\zeta)\,\mathrm{d}\zeta}{q_+(\zeta)-q}\log\left[4i\sqrt{1-\frac{\zeta^2}{a^2}}\sqrt{\frac{\zeta^2}{b^2}-1}-(2-\frac{\zeta^2}{b^2})^2\right]\,.
\end{eqnarray}
After a change of variable to $J=\zeta^2$ one has
\begin{equation}
\log T^+ = -\frac{1}{2\pi i}\int_{b^2}^{a^2}\! \frac{q_+ '(J)\,\mathrm{d}J}{q_+(J)-q}\lbrace \log(1-iz)-\log(1+iz)\rbrace\,,
\end{equation}
where $z(J) = 4\sqrt{1-\frac{J}{a^2}}\sqrt{\frac{J}{b^2}-1}/\left(2-\frac{J}{b^2}\right)^2$. Using trigonometric identities, one can simplify  $\log T^+$ further to obtain
\begin{equation}
\log T^+  =\frac{1}{\pi}\int_{b^2}^{a^2}\! \frac{q_+ '(J)\,\mathrm{d}J}{q_+(J)-q}\arctan\left[4\frac{\sqrt{1-\frac{J}{a^2}}\sqrt{\frac{J}{b^2}-1}}{(2-\frac{J}{b^2})^2}\right]\,.
\end{equation}
A similar calculation shows that 
\begin{equation}
\log T^- = -\frac{1}{\pi}\int_{b^2}^{a^2}\! \frac{q_- '(J)\,\mathrm{d}J}{q_-(J)-q}\arctan\left[4\frac{\sqrt{1-\frac{J}{a^2}}\sqrt{\frac{J}{b^2}-1}}{(2-\frac{J}{b^2})^2}\right]\
\end{equation}

We can now develop $T^\pm$ in powers of $1/(iq)$.
Denoting
\begin{equation}
\Pi^\pm_n = \mp\frac{1}{\pi}\int_{b^2}^{a^2}i^n q_\pm '(J) q_\pm(J)^{n-1}\arctan[z(J)]\,\mathrm{d}J\,
\end{equation}
we find that,
\begin{equation}
\log T^\pm = \sum_{n=1}^\infty \Pi^\pm_n \frac{1}{(iq)^n}\,.
\end{equation}
Therefore up to the 2\textsuperscript{nd}-order,
\begin{equation}
T^{\pm} = \exp\left(\sum_{n=1}^\infty \Pi^\pm_n \frac{1}{(iq)^n}\right)\simeq 1+\frac{\Pi_1^{\pm}}{iq}+\left(\Pi^\pm_2+\frac{1}{2}(\Pi^\pm_1)^2\right)\frac{1}{(iq)^2}\,.
\end{equation}
Also, note that 
\begin{equation}
\left(\Pi^+_n\right)^*=(-1)^{n+1}\Pi^-_n\,,
\end{equation}
and that
\begin{equation}
\Pi^\pm_n = -\frac{ i^{n-1}}{2\pi}\oint_{C_\pm} \xi^{n-1} \log T \!\mathrm{d}\xi\,\,.
\end{equation}
We can now substitute the expansions of $T^\pm$ back in Eqs.~(\ref{Upsilons}).
Equating
\begin{equation}
\widehat{\Upsilon}^+ = \frac{1}{(2i)^{1/2}} K_0^\sigma\frac{q-q_c}{\sqrt{q-q_a}}T^+ = \frac{1}{(2i)^{1/2}} K_0^\sigma q^{1/2}\left(1+\sum_n \frac{\hat{\Lambda}_n^+}{(iq)^n}\right) \,
\end{equation}
where $q_c =-\frac{V\omega\gamma_c^2}{c^2}+iQ_c$. Since
\begin{equation}
\frac{q-q_c}{\sqrt{q-q_a}} = q^{1/2}\left(1+\frac{iq_a/2-iq_c}{iq}+\frac{(1/2)q_aq_c-(3/8)q_a^2}{(iq)^2} + O(iq)^3\right)
\end{equation}
we obtain
\begin{eqnarray}
\hat{\Lambda}_1^+ &=& i\frac{q_a}{2}-iq_c+\Pi_1^+ \nonumber \\
\hat{\Lambda}_2^+ &=& \frac{1}{2} q_a q_c -\frac{3}{8}q_a^2+\Pi_2^+ +\frac{1}{2}(\Pi_1^+)^2 +\Pi_1^+ \left(i\frac{q_a}{2}-iq_c\right)\,.
\end{eqnarray}
The first of these equations corresponds to Eq. (A.25) in \cite{RamanathanThesis}
and the second corresponds to Eq. (A.35) in the limit $V=0$.

Now considering
\begin{equation}
\frac{1}{\widehat{\Upsilon}^-} = \frac{i^{3/2}(1-\nu^2) K_0^u}{2^{1/2} E} \frac{\sqrt{q-q_a^*}}{q-q_c^*}\frac{1}{T^-}  
= \frac{i^{3/2}(1-\nu^2) K_0^u}{2^{1/2} E}q^{-1/2}\left(1+\sum_n \frac{\hat{\Lambda}_n^-}{(iq)^n}\right) \,
\end{equation}
we develop in the same way 
\begin{equation}
\frac{\sqrt{q-q_a^*}}{q-q_c^*} = q^{-1/2}\left(1+\frac{iq_c^*-iq_a^*/2}{iq}+\frac{q_a^*q_c^*/2+(q_a^*)^2/8-(q_c^*)^2}{(iq)^2}\right)
\end{equation}
and therefore
\begin{eqnarray}
\hat{\Lambda}_1^- &=& iq_c^*-i\frac{q_a^*}{2}+\Pi_1^- \nonumber \\
\hat{\Lambda}_2^- &=& \frac{1}{2} q_a^* q_c^*+ \frac{1}{8}(q_a^*)^2-(q_c^*)^2+\Pi_2^- +\frac{1}{2}(\Pi_1^-)^2 +\frac{1}{2}\Pi_1^- \left(iq_c^*-i\frac{q_a^*}{2}\right)\,.
\end{eqnarray}

\subsection{Evaluation of the energy-release-rate}

To compute $G$ we can now use Eq.~(\ref{eq:G_equal_KuKs}) and substitute Eqs.~(\ref{eq:Ku_expansion},\ref{eq:Ks_expansion}). Since explicit expressions for $\Lambda_n^\pm$ were obtained in $(k,\omega)$ space, we will compute the Fourier transform $\widehat{G}(k,\omega)$. 
First, note that from Eqs.~(\ref{eq:Ku_expansion},\ref{eq:Ks_expansion})
\begin{equation}
\begin{split}
\frac{\widehat{K}^{\sigma,u}(k,\omega)}{K^{\sigma,u}_0} = (2\pi)^2\delta(k)\delta(\omega) - \hat{\Lambda_1}^\pm \hat{f} +\hat{\Lambda}_1^\pm\lbrace \hat{f}\otimes\left(\hat{\Lambda}_1^\pm \hat{f}\right)\rbrace \\
-\frac{1}{2}(\hat{\Lambda}_1^\pm)^2 (\hat{f}\otimes \hat{f}) -\hat{f}\otimes\left(\hat{\Lambda}_2^\pm \hat{f}\right)+\frac{1}{2}\hat{\Lambda}_2^\pm(\hat{f}\otimes \hat{f})\,,
\end{split}
\end{equation}
where
\begin{equation}
\widehat{(f\cdot g)}(k,\omega) = \widehat{f}\otimes\widehat{g} = \int\frac{d\omega'}{2\pi}\frac{dk'}{2\pi}\hat{f}(k-k',\omega-\omega')\hat{g}(k',\omega')\,.
\end{equation}
Then, the Fourier transformed Eq.~(\ref{eq:G_equal_KuKs}) becomes
\begin{equation}
\begin{split}
\frac{\widehat{G}(k,\omega)}{G_rg(V)}= (2\pi)^2\delta(k)\delta(\omega) - (\hat{\Lambda}_1^+ + \hat{\Lambda}_1^-) \hat{f}   +\hat{\Lambda}_1^+ \lbrace \hat{f}\otimes\left(\hat{\Lambda}_1^+ \hat{f}\right)\rbrace 
+ \hat{\Lambda}_1^- \lbrace \hat{f}\otimes\left(\hat{\Lambda}_1^-  \hat{f}\right)\rbrace \\ +\frac{1}{2}\left(\hat{\Lambda}_2^+ +  \hat{\Lambda}_2^- -(\hat{\Lambda}_1^+)^2 -(\hat{\Lambda}_1^-)^2\right) (\hat{f}\otimes \hat{f}) -\hat{f}\otimes\lbrace(\hat{\Lambda}_2^+ + \hat{\Lambda}_2^-) \hat{f}\rbrace  +(\hat{\Lambda}_1^+ \hat{f})\otimes(\hat{\Lambda}_1^- \hat{f})\,.
\end{split}
\end{equation}
Let us write explicit expressions for each of these terms.  First,
using the expressions of $\hat{\Lambda}_1^+$ and $\hat{\Lambda}_1^-$, we obtain
\begin{equation}
\widehat{\delta G^{(1)}}(k,\omega)=\hat{\Lambda}_1^+ + \hat{\Lambda}_1^- = 2|k|P_1\left(\frac{\omega}{|k|};V,\nu\right)\,,
\end{equation}
where
\begin{equation}
\begin{split}
P_1(s;V,\nu) &= -\frac{1}{2}\gamma_a\sqrt{1-\frac{\gamma_a^2 s^2}{a^2}}+\gamma_c\sqrt{1-\frac{\gamma_c^2 s^2}{c^2}}\\&+\frac{1}{2\pi}\int_{b^2}^{a^2}\frac{(s^2+V^2) (J+V^2)-2 J V^2}{ \sqrt{J} \sqrt{J-(s^2+V^2)} \left(J-V^2\right)^2}  \arctan[z(J)]\,\mathrm{d}J\,.
\end{split}
\end{equation}
The branch cuts of the square roots are defined so that $\sqrt{1-s^2}=i\,\mathrm{sign}(s)\sqrt{s^2-1}$ for $|s|>1$. This definition is consistent with the physical demand that elastic waves are radiated \textit{away} from the crack front~\citep{geubelleSpectralMethodThreedimensional1995,RamanathanThesis}.

Second, separating real and imaginary parts of the term $\hat{\Lambda}_1^+ \lbrace \hat{f}\otimes\left(\hat{\Lambda}_1^+ \hat{f}\right)\rbrace 
+ \hat{\Lambda}_1^- \lbrace \hat{f}\otimes\left(\hat{\Lambda}_1^-  \hat{f}\right)\rbrace$ yields
\begin{equation}
\begin{split}
&\hat{\Lambda}_1^+ \lbrace \hat{f}\otimes\left(\hat{\Lambda}_1^+ \hat{f}\right)\rbrace 
+ \hat{\Lambda}_1^- \lbrace \hat{f}\otimes\left(\hat{\Lambda}_1^-  \hat{f}\right)\rbrace  = \\ & 2\mathrm{Re}(\hat{\Lambda}_1^+) \lbrace \hat{f}\otimes\left(\mathrm{Re}(\hat{\Lambda}_1)^+ \hat{f}\right)\rbrace 
- 2\mathrm{Im}(\hat{\Lambda}_1^+) \lbrace \hat{f}\otimes\left(Im(\hat{\Lambda}_1^+) \hat{f}\right)\rbrace\,.
\end{split}
\end{equation}
Notice that the imaginary part of $\hat{\Lambda}_1$ is  linear in $\omega$
\begin{eqnarray}
\mathrm{Im}(\hat{\Lambda}^+_1) = V  Q_1(V) \omega \,,
\end{eqnarray}
with 
\begin{equation}
Q_1(V)  = \frac{1}{2(a^2-V^2)}-\frac{ 1}{c^2-V^2}+\frac{1}{\pi}\int_{b^2}^{a^2}\frac{1}{  \left(J-V^2\right)^2}  \arctan[z(J)]\,\mathrm{d}J
\,,
\end{equation}
or~\citep{adda-bediaDynamicStabilityCrack2013,Freund.90}
\begin{equation}
Q_1(V)  = -\frac{1}{2V}\frac{\mathrm{d}}{\mathrm{d}V}\left(\log\left(\frac{V^2}{R(V)\gamma_a b^2}\right)\right)\,,
\end{equation}
where $R(V) = 4\sqrt{1-\frac{V^2}{a^2 }}\sqrt{1-\frac{V^2}{b^2 }}-\left(2-\frac{V^2}{b^2 }\right)^2$. Hence,
\begin{equation}
\hat{\Lambda}_1^+ \lbrace \hat{f}\otimes\left(\hat{\Lambda}_1^+ \hat{f}\right)\rbrace 
+ \hat{\Lambda}_1^- \lbrace \hat{f}\otimes\left(\hat{\Lambda}_1^-  \hat{f}\right)\rbrace  =  2|k|P_1 \lbrace \hat{f}\otimes\left(|k|P_1 \hat{f}\right)\rbrace 
- 2 V^2 Q_1(V)^2 \omega \lbrace \hat{f}\otimes\left(\omega  \hat{f}\right)\rbrace\,.
\end{equation}

Third, to simplify the terms $\frac{1}{2}\left(\hat{\Lambda}_2^+ +  \hat{\Lambda}_2^- -(\hat{\Lambda}_1^+)^2 -(\hat{\Lambda}_1^-)^2\right) (\hat{f}\otimes \hat{f})$ and $f\otimes\lbrace(\hat{\Lambda}_2^+ + \hat{\Lambda}_2^-) \hat{f}\rbrace$,
 we notice that
\begin{equation}
\hat{\Lambda}_2^+ = \frac{1}{2}(\hat{\Lambda}^+_1)^2+\Pi_2^++\frac{1}{2}\left(q_c^2-\frac{1}{2}q_a^2 \right),\,\hat{\Lambda}_2^- = \frac{1}{2}(\hat{\Lambda}^-_1)^2+\Pi_2^--\frac{1}{2}\left(q_c^2-\frac{1}{2}q_a^2\right)^*
\end{equation}
Since $\hat{\Lambda}^-_1=(\hat{\Lambda}^+_1)^*$ and $(\Pi^+_2)^*=-\Pi^-_2$, we find 
\begin{equation}
\hat{\Lambda}_2^+ +  \hat{\Lambda}_2^- =   \Re[(\hat{\Lambda}_1^+)^2]+2i\Im[\Pi_2^+]+i\Im[q_c^2-\frac{1}{2}q_a^2]\,.
\end{equation}
Using the identity $\Re[(\hat{\Lambda}_1^+)^2] = \Re[\hat{\Lambda}_1^+]^2-\Im[\hat{\Lambda}_1^+]^2=k^2 P_1^2-V^2Q_1^2\omega^2$ and defining $iV\omega|k| P_2(\omega/|k|;V,\nu) = -2i\Im[\Pi_2^+]-i\Im[q_c^2-\frac{1}{2}q_a^2]$
we write
\begin{equation}\label{sum_Lambda_2}
\hat{\Lambda}_2^+ +  \hat{\Lambda}_2^- =   k^2 P_1^2-V^2Q_1^2\omega^2-iV\omega|k| P_2\,.
\end{equation}
Thus, we can write
\begin{equation}
\frac{1}{2}\left(\hat{\Lambda}_2^+ +  \hat{\Lambda}_2^- -(\hat{\Lambda}_1^+)^2 -(\hat{\Lambda}_1^-)^2\right) = -\frac{1}{2}\left(k^2 P_1^2-V^2Q_1^2\omega^2\right)-\frac{i}{2}V\omega|k|P_2\,,
\end{equation}
where
\begin{equation}
\begin{split}
P_2(s;V,\nu) = &2\frac{\gamma_c^3}{c^2}\sqrt{1-\frac{\gamma_c^2s^2}{c^2}}-\frac{\gamma_a^3}{a^2}\sqrt{1-\frac{\gamma_a^2s^2}{a^2}}\\
&+\frac{2}{\pi}\int_{b^2}^{a^2}\frac{(s^2+V^2) (3J+V^2)-2 J(J+ V^2)}{ \sqrt{J-(s^2+V^2)} \left(J-V^2\right)^3}  \arctan[z(J)]\,\frac{\mathrm{d}J}{2\sqrt{J}}\,.
\end{split}
\end{equation}
Finally, using the distributive law, one obtains\begin{equation}
(\hat{\Lambda}_1^+ \hat{f})\otimes(\hat{\Lambda}_1^- \hat{f}) = (|k|P_1 \hat{f})\otimes(|k|P_1 \hat{f})+V^2Q_1^2(\omega \hat{f})\otimes(\omega \hat{f})\,.
\end{equation}

Collecting all the above expressions, the 2\textsuperscript{nd}-order correction to the energy-release-rate can be written as
\begin{equation}
\begin{split}
& \widehat{\delta G^{(2)}}(k,\omega) = \\ & 2|k|P_1\lbrace \hat{f}\otimes\left(|k|P_1 \hat{f}\right)\rbrace 
- 2 Q_1^2 V^2\omega \lbrace \hat{f}\otimes\left(\omega  \hat{f}\right)\rbrace -\left\{ \frac{1}{2}\left(k^2 P_1^2-V^2Q_1^2\omega^2\right)+\frac{i}{2}V\omega|k|P_2\right\}(\hat{f}\otimes \hat{f})\\& -\hat{f}\otimes\left\{ \left(k^2 P_1^2-V^2Q_1^2\omega^2-iV\omega|k| P_2\right) \hat{f}\right\}   +(|k|P_1 \hat{f})\otimes(|k|P_1 \hat{f})+V^2 Q_1^2 (\omega \hat{f})\otimes(\omega \hat{f})\,.
\end{split}
\end{equation}
This expression can be further simplified using the identity $\hat{f}\otimes(\omega \hat{f})= \frac{1}{2} \omega (\hat{f}\otimes \hat{f})$.
Then,
\begin{equation}\label{G2_final}
\begin{split}
\widehat{\delta G^{(2)}}(k,\omega)=
& 2|k|P_1\lbrace \hat{f}\otimes\left(|k|P_1 \hat{f}\right)\rbrace 
-\left\{ \frac{1}{2}k^2 P_1^2+\frac{i}{2}V\omega|k|P_2\right\}(\hat{f}\otimes \hat{f})\\& -\hat{f}\otimes\left\{ \left(k^2 P_1^2-iV\omega|k| P_2\right)\hat{f}\right\}   +(|k|P_1 \hat{f} )\otimes(|k|P_1 \hat{f})\,.
\end{split}
\end{equation}

\subsection{Transformation to \texorpdfstring{$(k,t)$}{(k,t)} space}

To obtain a real space expression of the form $G = g(V_\perp)( 1+ H^{(1)}[f] + H^{(2)}[f,f] +\order{f^3})$ where the functionals $H^{(i)}$ only depend on the history of the front dynamics$\left\{f(z,t');\; t'<t\right\}$, we transform $\widehat{\delta G^{(1)}}(k,\omega)$ and $\widehat{
\delta G^{(2)}}(k,\omega)$ from $\omega$ to $t$. Following~\cite{Morrissey.00}, we use the Bessel identities
$$
    \frac{1}{2\pi}\int\!\mathrm{d}\omega\,e^{i\omega t}  \sqrt{\beta^2-\omega^2} = \delta'(t) + p^2 \frac{J_1(pt)}{pt}\mathcal{H}(t)
$$

and

$$
    \frac{1}{2\pi}\int\!\mathrm{d}\omega\,e^{i\omega t}  \frac{1}{\sqrt{\beta^2-\omega^2}} =  J_0(pt)\mathcal{H}(t)
$$
where $\mathcal{H}(t)$ is the Heaviside function, to write
\begin{equation}
\begin{split}
    &\frac{1}{2\pi}\int\!\mathrm{d}\omega\,e^{i\omega t}  |k|P_1\left(\frac{\omega}{|k|}\right) = \left(-\frac{1}{2}
    \frac{a}{a^2-V^2} + \frac{c}{c^2-V^2}\right)\delta'(t)\\
    &-\partial_t^2\frac{1}{2} \int_b^a\!\mathrm{d}\eta\,\Theta(\eta)\frac{\eta^2+V^2}{(\eta^2-V^2)^2}J_0(\beta_\eta t )\mathcal{H}(t)\\
    &+k^2 \left(-\frac{a}{2}\frac{J_1(\beta_a t)}{\beta_a t} + c\frac{J_1(\beta_c t)}{\beta_c t}-\frac{1}{2}\int_b^a\!\mathrm{d}\eta\,\Theta(\eta)\frac{V^2}{\eta^2-V^2}J_0(\beta_\eta t )\right)\mathcal{H}(t)    
\end{split}
\end{equation}
where $\beta_s = \sqrt{s^2-V^2}|k|$. From the Bessel identities $J_0'(x) = - J_1(x)$ and $J_0''(x) =  (J_2(x)-J_0(x))/2$

\begin{equation}
    \frac{1}{2\pi}\int\!\mathrm{d}\omega\,e^{i\omega t}  |k|P_1\left(\frac{\omega}{|k|}\right) = C_1\delta'(t) +B_1(k,t)\delta(t)  + A_1(k,t)\mathcal{H}(t)
\end{equation}

\begin{eqnarray}
    C_1(k,t) &=& -\frac{1}{2}\frac{a}{a^2-V^2} + \frac{c}{c^2-V^2} - \frac{1}{2}\int_b^a\!\mathrm{d}\eta\,\Theta(\eta)\frac{\eta^2+V^2}{(\eta^2-V^2)^{2}}\ J_0(\beta_\eta t )\nonumber\\ 
    B_1(k,t) &=& -|k|\int_b^a\!\mathrm{d}\eta\,\Theta(\eta)\frac{\eta^2+V^2}{(\eta^2-V^2)^{3/2}}\ J_1(\beta_\eta t )\nonumber\\
    A_1(k,t) &=& k^2\left[-\frac{a}{2} \frac{J_1(\beta_a t)}{\beta_a t} + c \frac{J_1(\beta_c t)}{\beta_c t} - \frac{1}{4} \int_b^a\!\mathrm{d}\eta\,\Theta(\eta)\left(\frac{\eta^2+V^2}{\eta^2-V^2}J_2(\beta_\eta t)-J_0(\beta_\eta t)\right)\right]\,.\nonumber
\end{eqnarray}

Then,
\begin{equation}
    \frac{1}{2\pi}\int\!\mathrm{d}\omega\, e^{i\omega t} |k|P_1\left(\frac{\omega}{|k|}\right)\hat{f}(k,\omega) = -\pi_1 \bar{f}_{t} - I_1[\bar{f}]
\end{equation}

where $\pi_1$ is given by Eq.~(\ref{section2_eq_pi1}), and $I_1[\bar{f}] = -\int_{-\infty}^t\!\mathrm{d}t'\,A_1(k,t-t')\bar{f}(k,t')$.

Since $2\pi_1 =g'(V)/g(V)$, to the 1\textsuperscript{st}-order in $f$ we have 
$G =  g(V_\perp)(1 + H^{(1)}[f]+\order{f^2})$  where $\overline{H^{(1)}[f]} = 2I_1[\bar{f}]$~\citep{Morrissey.00}.

To extend this result to the 2\textsuperscript{nd}-order, we use the Bessel identity $J_0'''(x) = 3J_1(x)/4-J_3(x)/4$ to find
\begin{equation}
    \frac{1}{2\pi}\int\!\mathrm{d}\omega\, e^{i\omega t} i\omega|k|P_2\left(\frac{\omega}{|k|}\right) = D_2\delta''(t) +C_2\delta'(t) + B_2\delta(t) +A_2\mathcal{H}(t)\,,
\end{equation}
where
\begin{eqnarray}
    D_2(k,t) &=&2 \frac{c}{(c^2-V^2)^2} - \frac{a}{(a^2-V^2)^2} -\int_b^a\!\mathrm{d}\eta\,\Theta(\eta)\frac{3\eta^2+V^2}{(\eta^2-V^2)^{3}} J_0(\beta_\eta t)\nonumber\\\nonumber
    C_2(k,t) &=& 3|k|\int_b^a\!\mathrm{d}\eta\,\Theta(\eta)\frac{3\eta^2+V^2}{(\eta^2-V^2)^{5/2}} J_1(\beta_\eta t)\\\nonumber
    B_2(k,t) &=&k^2\Bigg[2\frac{c}{c^2-V^2}\frac{J_1(\beta_c t)}{\beta_c t} - \frac{a}{a^2-V^2}\frac{J_1(\beta_a t)}{\beta_a t}\\\nonumber &+&\frac{1}{2}\int_b^a\!\mathrm{d}\eta\,\frac{\Theta(\eta)}{(\eta^2-V^2)^2}\left((5\eta^2+V^2) J_0(\beta_\eta t)-(9\eta^2+3V^2)J_2(\beta_\eta t)\right)\Bigg]\\\nonumber
    A_2(k,t) &=& |k|^3\Bigg[-\gamma_a \frac{J_2(\beta_a t)}{\beta_a t} + 2\gamma_c \frac{J_2(\beta_c t)}{\beta_c t}\\\nonumber &-&\frac{1}{4} \int_b^a\!\mathrm{d}\eta\,\frac{\Theta(\eta)}{\sqrt{\eta^2-V^2}}\left(\frac{3\eta^2+V^2}{\eta^2-V^2}J_3(\beta_\eta t)-J_1(\beta_\eta t)\right)\Bigg]\,.\nonumber
\end{eqnarray}

Then,
\begin{equation}
    \frac{1}{2\pi}\int\!\mathrm{d}\omega\, e^{i\omega t} i\omega|k|P_2\left(\frac{\omega}{|k|}\right)\hat{f}(k,\omega) = -\pi_2 \bar{f}_{tt} -\pi_1 k^2 \bar{f} - I_2[\bar{f}]
\end{equation}
where 
\begin{equation}\label{section2_eq_pi2}
\pi_2 = \frac{a}{(a^2-V^2)^2} -2\frac{c}{(c^2-V^2)^2} +\int_b^a \frac{3\eta^2+V^2}{(\eta^2-V^2)^3}\Theta(\eta)\mathrm{d}\eta\,.
\end{equation}
The second term results from the identity $\partial_t^2 D_2(0) - \partial_t C_2(0)+B_2(0)= -\pi_1 k^2$, and the third term is $I_2[\bar{f}] = -\int_{-\infty}^t\!\mathrm{d}t \, A_2(k,t-t')\bar{f}(k,t')$.

The 2\textsuperscript{nd}-order contribution to $G$ is then,
\begin{equation}
    \begin{split}
        \overline{\delta G^{(2)}}(k,t) &= (2\pi_1^2+V\pi_2)\bar{f}_t*\bar{f}_t + 2\pi_1 \bar{f}_t*\overline{H^{(1)}[f]}+ V\pi_1(k\bar{f})*(k\bar{f}) + \overline{H^{(2)}[f,f]}\,,
    \end{split}
\end{equation}
where the convolution operator is defined as $f*g = \int \!\mathrm{d}k'\,f(k-k')g(k')$ and
\begin{equation}
\begin{split}
\overline{H^{(2)}[f,f]} &= 2 I_1[\bar{f}*I_1[\bar{f}]]-\frac{1}{2}I_1[I_1[\bar{f}*\bar{f}]]-\bar{f}*I_1[I_1[\bar{f}]]+I_1[\bar{f}]*I_1[\bar{f}]\nonumber\\
        &+\frac{V}{2}I_2[\bar{f}*\bar{f}] - V \bar{f}*I_2[\bar{f}]\nonumber\,.
\end{split}
\end{equation}

Using the identities $\pi_1'(V) = V\pi_2$ and $g''(V)/2g(V) = 2\pi_1^2+\pi_1'(V)$ we obtain
\begin{eqnarray}
    G(z,t) &=& G_r g(V)\Bigg(1 + \frac{g'(V)}{g(V)}f_t+\frac{g''(V)}{2g(V)}f_{t}^2-\frac{g'(V)}{g(V)}\frac{V}{2} f_z^2 \nonumber\\
    &+&\frac{g'(V)}{g(V)}f_t H^{(1)}[f]+ H^{(1)}[f]+H^{(2)}[f,f]+\order{f^3}\Bigg)\,,
\end{eqnarray}
which is equivalent to
\begin{equation}
    G(z,t) = G_r g(V_\perp)\left(1 + H^{(1)}[f]+H^{(2)}[f,f]+\order{f^3}\right)\,.
\end{equation}

\section*{CRediT authorship contribution statement}

Itamar Kolvin: Conceptualization, Methodology,  Validation, Investigation, Writing, Visualization. Mokhtar Adda-Bedia: Conceptualization, Methodology, Investigation, Writing, Funding Acquisition.

\section*{Declaration of Competing Interest}

The authors declare that they have no known competing financial interests or personal relationships that could have appeared to influence the work reported in this paper.

\section*{Acknowledgments}
This work was supported by the International Research Project ``Non-Equilibrium Physics of Complex Systems'' (IRP-PhyComSys, France-Israel).

\bibliographystyle{elsarticle-harv}
\bibliography{CFWT}

\begin{thebibliography}{64}
\expandafter\ifx\csname natexlab\endcsname\relax\def\natexlab#1{#1}\fi
\providecommand{\url}[1]{\texttt{#1}}
\providecommand{\href}[2]{#2}
\providecommand{\path}[1]{#1}
\providecommand{\DOIprefix}{doi:}
\providecommand{\ArXivprefix}{arXiv:}
\providecommand{\URLprefix}{URL: }
\providecommand{\Pubmedprefix}{pmid:}
\providecommand{\doi}[1]{\href{http://dx.doi.org/#1}{\path{#1}}}
\providecommand{\Pubmed}[1]{\href{pmid:#1}{\path{#1}}}
\providecommand{\bibinfo}[2]{#2}
\ifx\xfnm\relax \def\xfnm[#1]{\unskip,\space#1}\fi
\bibitem[{{Adda-Bedia} et~al.(2013){Adda-Bedia}, Arias, Bouchbinder and
  Katzav}]{adda-bediaDynamicStabilityCrack2013}
\bibinfo{author}{{Adda-Bedia}, M.}, \bibinfo{author}{Arias, R.E.},
  \bibinfo{author}{Bouchbinder, E.}, \bibinfo{author}{Katzav, E.},
  \bibinfo{year}{2013}.
\newblock \bibinfo{title}{Dynamic {{Stability}} of {{Crack Fronts}}:
  {{Out-Of-Plane Corrugations}}}.
\newblock \bibinfo{journal}{Physical Review Letters} \bibinfo{volume}{110},
  \bibinfo{pages}{014302}.
\newblock \DOIprefix\doi{10.1103/PhysRevLett.110.014302}.
\bibitem[{{Adda-Bedia} et~al.(2006){Adda-Bedia}, Katzav and
  Vandembroucq}]{Adda-Bedia.06}
\bibinfo{author}{{Adda-Bedia}, M.}, \bibinfo{author}{Katzav, E.},
  \bibinfo{author}{Vandembroucq, D.}, \bibinfo{year}{2006}.
\newblock \bibinfo{title}{Second-order variation in elastic fields of a tensile
  planar crack with a curved front}.
\newblock \bibinfo{journal}{Physical Review E} \bibinfo{volume}{73},
  \bibinfo{pages}{35106}.
\bibitem[{Albertini et~al.(2021)Albertini, Lebihain, Hild, Ponson and
  Kammer}]{albertiniEffectiveToughnessHeterogeneous2021}
\bibinfo{author}{Albertini, G.}, \bibinfo{author}{Lebihain, M.},
  \bibinfo{author}{Hild, F.}, \bibinfo{author}{Ponson, L.},
  \bibinfo{author}{Kammer, D.S.}, \bibinfo{year}{2021}.
\newblock \bibinfo{title}{Effective {{Toughness}} of {{Heterogeneous
  Materials}} with {{Rate-Dependent Fracture Energy}}}.
\newblock \bibinfo{journal}{Physical Review Letters} \bibinfo{volume}{127},
  \bibinfo{pages}{035501}.
\newblock \DOIprefix\doi{10.1103/PhysRevLett.127.035501}.
\bibitem[{Amestoy et~al.(1981)Amestoy, Bui and
  Labbens}]{amestoyDefinitionLocalPath1981}
\bibinfo{author}{Amestoy, M.}, \bibinfo{author}{Bui, H.D.},
  \bibinfo{author}{Labbens, R.}, \bibinfo{year}{1981}.
\newblock \bibinfo{title}{On the definition of local path independent integrals
  in three-dimensional crack problems}.
\newblock \bibinfo{journal}{Mechanics Research Communications}
  \bibinfo{volume}{8}, \bibinfo{pages}{231--236}.
\newblock \DOIprefix\doi{10.1016/0093-6413(81)90058-6}.
\bibitem[{Bayart et~al.(2018)Bayart, Svetlizky and
  Fineberg}]{bayartRuptureDynamicsHeterogeneous2018}
\bibinfo{author}{Bayart, E.}, \bibinfo{author}{Svetlizky, I.},
  \bibinfo{author}{Fineberg, J.}, \bibinfo{year}{2018}.
\newblock \bibinfo{title}{Rupture {{Dynamics}} of {{Heterogeneous Frictional
  Interfaces}}}.
\newblock \bibinfo{journal}{Journal of Geophysical Research: Solid Earth}
  \bibinfo{volume}{123}, \bibinfo{pages}{3828--3848}.
\newblock \DOIprefix\doi{10.1002/2018JB015509}.
\bibitem[{Bedford et~al.(2022)Bedford, Faulkner and
  Lapusta}]{bedfordFaultRockHeterogeneity2022}
\bibinfo{author}{Bedford, J.D.}, \bibinfo{author}{Faulkner, D.R.},
  \bibinfo{author}{Lapusta, N.}, \bibinfo{year}{2022}.
\newblock \bibinfo{title}{Fault rock heterogeneity can produce fault weakness
  and reduce fault stability}.
\newblock \bibinfo{journal}{Nature Communications} \bibinfo{volume}{13},
  \bibinfo{pages}{326}.
\newblock \DOIprefix\doi{10.1038/s41467-022-27998-2}.
\bibitem[{Bleyer and
  Molinari(2017)}]{bleyerMicrobranchingInstabilityPhasefield2017}
\bibinfo{author}{Bleyer, J.}, \bibinfo{author}{Molinari, J.F.},
  \bibinfo{year}{2017}.
\newblock \bibinfo{title}{Microbranching instability in phase-field modelling
  of dynamic brittle fracture}.
\newblock \bibinfo{journal}{Applied Physics Letters} \bibinfo{volume}{110},
  \bibinfo{pages}{151903}.
\bibitem[{Buehler(2022)}]{buehlerModelingAtomisticDynamic2022}
\bibinfo{author}{Buehler, M.J.}, \bibinfo{year}{2022}.
\newblock \bibinfo{title}{Modeling {{Atomistic Dynamic Fracture Mechanisms
  Using}} a {{Progressive Transformer Diffusion Model}}}.
\newblock \bibinfo{journal}{Journal of Applied Mechanics} \bibinfo{volume}{89}.
\newblock \DOIprefix\doi{10.1115/1.4055730}.
\bibitem[{Chen et~al.(2015)Chen, Cambonie, Lazarus, Nicoli, Pons and
  Karma}]{Chen.15}
\bibinfo{author}{Chen, C.H.}, \bibinfo{author}{Cambonie, T.},
  \bibinfo{author}{Lazarus, V.}, \bibinfo{author}{Nicoli, M.},
  \bibinfo{author}{Pons, A.J.}, \bibinfo{author}{Karma, A.},
  \bibinfo{year}{2015}.
\newblock \bibinfo{title}{Crack front segmentation and facet coarsening in
  mixed-mode fracture}.
\newblock \bibinfo{journal}{Physical review letters} \bibinfo{volume}{115},
  \bibinfo{pages}{265503}.
\bibitem[{Chopin et~al.(2015)Chopin, Boudaoud and
  {Adda-Bedia}}]{chopinMorphologyDynamicsCrack2015}
\bibinfo{author}{Chopin, J.}, \bibinfo{author}{Boudaoud, A.},
  \bibinfo{author}{{Adda-Bedia}, M.}, \bibinfo{year}{2015}.
\newblock \bibinfo{title}{Morphology and dynamics of a crack front propagating
  in a model disordered material}.
\newblock \bibinfo{journal}{Journal of the Mechanics and Physics of Solids}
  \bibinfo{volume}{74}, \bibinfo{pages}{38--48}.
\newblock \DOIprefix\doi{10.1016/j.jmps.2014.10.001}.
\bibitem[{Cochard et~al.(2024)Cochard, Svetlizky, Albertini, Viesca,
  Rubinstein, Spaepen, Yuan, Denolle, Song, Xiao and
  Weitz}]{cochardPropagationExtendedFractures2024}
\bibinfo{author}{Cochard, T.}, \bibinfo{author}{Svetlizky, I.},
  \bibinfo{author}{Albertini, G.}, \bibinfo{author}{Viesca, R.C.},
  \bibinfo{author}{Rubinstein, S.M.}, \bibinfo{author}{Spaepen, F.},
  \bibinfo{author}{Yuan, C.}, \bibinfo{author}{Denolle, M.},
  \bibinfo{author}{Song, Y.Q.}, \bibinfo{author}{Xiao, L.},
  \bibinfo{author}{Weitz, D.A.}, \bibinfo{year}{2024}.
\newblock \bibinfo{title}{Propagation of extended fractures by local nucleation
  and rapid transverse expansion of crack-front distortion}.
\newblock \bibinfo{journal}{Nature Physics} \bibinfo{volume}{20},
  \bibinfo{pages}{660--665}.
\newblock \DOIprefix\doi{10.1038/s41567-023-02365-0}.
\bibitem[{D{\'e}mery et~al.(2014)D{\'e}mery, Rosso and
  Ponson}]{demeryMicrostructuralFeaturesEffective2014}
\bibinfo{author}{D{\'e}mery, V.}, \bibinfo{author}{Rosso, A.},
  \bibinfo{author}{Ponson, L.}, \bibinfo{year}{2014}.
\newblock \bibinfo{title}{From microstructural features to effective toughness
  in disordered brittle solids}.
\newblock \bibinfo{journal}{EPL (Europhysics Letters)} \bibinfo{volume}{105},
  \bibinfo{pages}{34003}.
\newblock \DOIprefix\doi{10.1209/0295-5075/105/34003}.
\bibitem[{Dodds et~al.(1988)Dodds, Carpenter and
  Sorem}]{doddsNumericalEvaluation3D1988}
\bibinfo{author}{Dodds, R.H.}, \bibinfo{author}{Carpenter, W.C.},
  \bibinfo{author}{Sorem, W.A.}, \bibinfo{year}{1988}.
\newblock \bibinfo{title}{Numerical evaluation of a 3-{{D J-integral}} and
  comparison with experimental results for a 3-{{Point}} bend specimen}.
\newblock \bibinfo{journal}{Engineering Fracture Mechanics}
  \bibinfo{volume}{29}, \bibinfo{pages}{275--285}.
\newblock \DOIprefix\doi{10.1016/0013-7944(88)90017-3}.
\bibitem[{El~Kabir et~al.(2018)El~Kabir, Dubois, Moutou~Pitti, Recho and
  Lapusta}]{elkabirNewAnalyticalGeneralization2018}
\bibinfo{author}{El~Kabir, S.}, \bibinfo{author}{Dubois, F.},
  \bibinfo{author}{Moutou~Pitti, R.}, \bibinfo{author}{Recho, N.},
  \bibinfo{author}{Lapusta, Y.}, \bibinfo{year}{2018}.
\newblock \bibinfo{title}{A new analytical generalization of the {{J}} and
  {{G-theta}} integrals for planar cracks in a three-dimensional medium}.
\newblock \bibinfo{journal}{Theoretical and Applied Fracture Mechanics}
  \bibinfo{volume}{94}, \bibinfo{pages}{101--109}.
\newblock \DOIprefix\doi{10.1016/j.tafmec.2018.01.004}.
\bibitem[{Eriksson(2002)}]{erikssonDomainIndependentIntegral2002}
\bibinfo{author}{Eriksson, K.}, \bibinfo{year}{2002}.
\newblock \bibinfo{title}{A domain independent integral expression for the
  crack extension force of a curved crack in three dimensions}.
\newblock \bibinfo{journal}{Journal of the Mechanics and Physics of Solids}
  \bibinfo{volume}{50}, \bibinfo{pages}{381--403}.
\newblock \DOIprefix\doi{10.1016/S0022-5096(01)00059-X}.
\bibitem[{Eshelby(1969)}]{eshelbyElasticFieldCrack1969}
\bibinfo{author}{Eshelby, J.D.}, \bibinfo{year}{1969}.
\newblock \bibinfo{title}{The elastic field of a crack extending non-uniformly
  under general anti-plane loading}.
\newblock \bibinfo{journal}{Journal of the Mechanics and Physics of Solids}
  \bibinfo{volume}{17}, \bibinfo{pages}{177--199}.
\bibitem[{Fineberg et~al.(2003)Fineberg, Sharon and Cohen}]{Fineberg.03}
\bibinfo{author}{Fineberg, J.}, \bibinfo{author}{Sharon, E.},
  \bibinfo{author}{Cohen, G.}, \bibinfo{year}{2003}.
\newblock \bibinfo{title}{Crack front waves in dynamic fracture}.
\newblock \bibinfo{journal}{International Journal of Fracture}
  \bibinfo{volume}{121}, \bibinfo{pages}{55--69}.
\bibitem[{Freund(1990)}]{Freund.90}
\bibinfo{author}{Freund, L.B.}, \bibinfo{year}{1990}.
\newblock \bibinfo{title}{Dynamic Fracture Mechanics}.
\newblock \bibinfo{publisher}{Cambridge University Press},
  \bibinfo{address}{Cambridge; New York}.
\bibitem[{Geubelle and Rice(1995)}]{geubelleSpectralMethodThreedimensional1995}
\bibinfo{author}{Geubelle, P.H.}, \bibinfo{author}{Rice, J.R.},
  \bibinfo{year}{1995}.
\newblock \bibinfo{title}{A spectral method for three-dimensional elastodynamic
  fracture problems}.
\newblock \bibinfo{journal}{Journal of the Mechanics and Physics of Solids}
  \bibinfo{volume}{43}, \bibinfo{pages}{1791--1824}.
\newblock \DOIprefix\doi{10.1016/0022-5096(95)00043-I}.
\bibitem[{Goldman et~al.(2010)Goldman, Livne and
  Fineberg}]{goldmanAcquisitionInertiaMoving2010}
\bibinfo{author}{Goldman, T.}, \bibinfo{author}{Livne, A.},
  \bibinfo{author}{Fineberg, J.}, \bibinfo{year}{2010}.
\newblock \bibinfo{title}{Acquisition of {{Inertia}} by a {{Moving Crack}}}.
\newblock \bibinfo{journal}{Physical Review Letters} \bibinfo{volume}{104},
  \bibinfo{pages}{114301}.
\newblock \DOIprefix\doi{10.1103/PhysRevLett.104.114301}.
\bibitem[{Goswami et~al.(2022)Goswami, Yin, Yu and
  Karniadakis}]{goswamiPhysicsinformedVariationalDeepONet2022}
\bibinfo{author}{Goswami, S.}, \bibinfo{author}{Yin, M.}, \bibinfo{author}{Yu,
  Y.}, \bibinfo{author}{Karniadakis, G.E.}, \bibinfo{year}{2022}.
\newblock \bibinfo{title}{A physics-informed variational {{DeepONet}} for
  predicting crack path in quasi-brittle materials}.
\newblock \bibinfo{journal}{Computer Methods in Applied Mechanics and
  Engineering} \bibinfo{volume}{391}, \bibinfo{pages}{114587}.
\newblock \DOIprefix\doi{10.1016/j.cma.2022.114587}.
\bibitem[{Gounon et~al.(2022)Gounon, Latour, Letort and
  El~Arem}]{gounonRuptureNucleationPeriodically2022}
\bibinfo{author}{Gounon, A.}, \bibinfo{author}{Latour, S.},
  \bibinfo{author}{Letort, J.}, \bibinfo{author}{El~Arem, S.},
  \bibinfo{year}{2022}.
\newblock \bibinfo{title}{Rupture {{Nucleation}} on a {{Periodically
  Heterogeneous Interface}}}.
\newblock \bibinfo{journal}{Geophysical Research Letters} \bibinfo{volume}{49},
  \bibinfo{pages}{e2021GL096816}.
\newblock \DOIprefix\doi{10.1029/2021GL096816}.
\bibitem[{Gupta et~al.()Gupta, Esmaeeli and
  Moini}]{guptaToughDuctileArchitected}
\bibinfo{author}{Gupta, S.}, \bibinfo{author}{Esmaeeli, H.S.},
  \bibinfo{author}{Moini, R.}, .
\newblock \bibinfo{title}{Tough and {{Ductile Architected Nacre-Like
  Cementitious Composites}}}.
\newblock \bibinfo{journal}{Advanced Functional Materials}
  \bibinfo{volume}{n/a}, \bibinfo{pages}{2313516}.
\newblock \DOIprefix\doi{10.1002/adfm.202313516}.
\bibitem[{Heizler and Kessler(2015)}]{heizlerMicrobranchingModeIFracture2015}
\bibinfo{author}{Heizler, S.I.}, \bibinfo{author}{Kessler, D.A.},
  \bibinfo{year}{2015}.
\newblock \bibinfo{title}{Microbranching in mode-{{I}} fracture using
  large-scale simulations of amorphous and perturbed-lattice models}.
\newblock \bibinfo{journal}{Physical Review E} \bibinfo{volume}{92},
  \bibinfo{pages}{12403}.
\bibitem[{Henry(2019)}]{henryLimitationsModellingCrack2019}
\bibinfo{author}{Henry, H.}, \bibinfo{year}{2019}.
\newblock \bibinfo{title}{Limitations of the modelling of crack propagating
  through heterogeneous material using a phase field approach}.
\newblock \bibinfo{journal}{Theoretical and Applied Fracture Mechanics}
  \bibinfo{volume}{104}, \bibinfo{pages}{102384}.
\newblock \DOIprefix\doi{10.1016/j.tafmec.2019.102384}.
\bibitem[{Henry and {Adda-Bedia}(2013)}]{henryFractographicAspectsCrack2013a}
\bibinfo{author}{Henry, H.}, \bibinfo{author}{{Adda-Bedia}, M.},
  \bibinfo{year}{2013}.
\newblock \bibinfo{title}{Fractographic aspects of crack branching instability
  using a phase-field model}.
\newblock \bibinfo{journal}{Physical Review E} \bibinfo{volume}{88},
  \bibinfo{pages}{060401}.
\newblock \DOIprefix\doi{10.1103/PhysRevE.88.060401}.
\bibitem[{Kolvin and {Adda-Bedia}(2024)}]{kolvinDualRoleHeterogeneity2024}
\bibinfo{author}{Kolvin, I.}, \bibinfo{author}{{Adda-Bedia}, M.},
  \bibinfo{year}{2024}.
\newblock \bibinfo{title}{Dual {{Role}} for {{Heterogeneity}} in {{Dynamic
  Fracture}}}.
\newblock \DOIprefix\doi{10.48550/arXiv.2407.02347},
  \href{http://arxiv.org/abs/2407.02347}{{\tt arXiv:2407.02347}}.
\bibitem[{Kolvin et~al.(2017)Kolvin, Fineberg and
  {Adda-Bedia}}]{kolvinNonlinearFocusingDynamic2017}
\bibinfo{author}{Kolvin, I.}, \bibinfo{author}{Fineberg, J.},
  \bibinfo{author}{{Adda-Bedia}, M.}, \bibinfo{year}{2017}.
\newblock \bibinfo{title}{Nonlinear {{Focusing}} in {{Dynamic Crack Fronts}}
  and the {{Microbranching Transition}}}.
\newblock \bibinfo{journal}{Physical review letters} \bibinfo{volume}{119},
  \bibinfo{pages}{215505}.
\bibitem[{Latour et~al.(2013)Latour, Voisin, Renard, Larose, Catheline and
  Campillo}]{latourEffectFaultHeterogeneity2013}
\bibinfo{author}{Latour, S.}, \bibinfo{author}{Voisin, C.},
  \bibinfo{author}{Renard, F.}, \bibinfo{author}{Larose, E.},
  \bibinfo{author}{Catheline, S.}, \bibinfo{author}{Campillo, M.},
  \bibinfo{year}{2013}.
\newblock \bibinfo{title}{Effect of fault heterogeneity on rupture dynamics:
  {{An}} experimental approach using ultrafast ultrasonic imaging}.
\newblock \bibinfo{journal}{Journal of Geophysical Research: Solid Earth}
  \bibinfo{volume}{118}, \bibinfo{pages}{5888--5902}.
\newblock \DOIprefix\doi{10.1002/2013JB010231}.
\bibitem[{Lebihain et~al.(2020)Lebihain, Leblond and
  Ponson}]{lebihainEffectiveToughnessPeriodic2020}
\bibinfo{author}{Lebihain, M.}, \bibinfo{author}{Leblond, J.B.},
  \bibinfo{author}{Ponson, L.}, \bibinfo{year}{2020}.
\newblock \bibinfo{title}{Effective toughness of periodic heterogeneous
  materials: The effect of out-of-plane excursions of cracks}.
\newblock \bibinfo{journal}{Journal of the Mechanics and Physics of Solids}
  \bibinfo{volume}{137}, \bibinfo{pages}{103876}.
\newblock \DOIprefix\doi{10.1016/j.jmps.2020.103876}.
\bibitem[{Lebihain et~al.(2021)Lebihain, Roch, Violay and
  Molinari}]{lebihainEarthquakeNucleationFaults2021}
\bibinfo{author}{Lebihain, M.}, \bibinfo{author}{Roch, T.},
  \bibinfo{author}{Violay, M.}, \bibinfo{author}{Molinari, J.F.},
  \bibinfo{year}{2021}.
\newblock \bibinfo{title}{Earthquake {{Nucleation Along Faults With
  Heterogeneous Weakening Rate}}}.
\newblock \bibinfo{journal}{Geophysical Research Letters} \bibinfo{volume}{48},
  \bibinfo{pages}{e2021GL094901}.
\newblock \DOIprefix\doi{10.1029/2021GL094901}.
\bibitem[{Leblond et~al.(2011)Leblond, Karma and Lazarus}]{Leblond.11}
\bibinfo{author}{Leblond, J.B.}, \bibinfo{author}{Karma, A.},
  \bibinfo{author}{Lazarus, V.}, \bibinfo{year}{2011}.
\newblock \bibinfo{title}{Theoretical analysis of crack front instability in
  mode {{I}}+\{\vphantom\}{{III}}\vphantom\{\}}.
\newblock \bibinfo{journal}{Journal of the Mechanics and Physics of Solids}
  \bibinfo{volume}{59}, \bibinfo{pages}{1872--1887}.
\newblock \DOIprefix\doi{10.1016/j.jmps.2011.05.011}.
\bibitem[{Leblond et~al.(2012)Leblond, Patinet, Frelat and
  Lazarus}]{leblondSecondorderCoplanarPerturbation2012}
\bibinfo{author}{Leblond, J.B.}, \bibinfo{author}{Patinet, S.},
  \bibinfo{author}{Frelat, J.}, \bibinfo{author}{Lazarus, V.},
  \bibinfo{year}{2012}.
\newblock \bibinfo{title}{Second-order coplanar perturbation of a semi-infinite
  crack in an infinite body}.
\newblock \bibinfo{journal}{Engineering Fracture Mechanics}
  \bibinfo{volume}{90}, \bibinfo{pages}{129--142}.
\bibitem[{Leguillon(2014)}]{leguillonAttemptExtend2D2014}
\bibinfo{author}{Leguillon, D.}, \bibinfo{year}{2014}.
\newblock \bibinfo{title}{An attempt to extend the {{2D}} coupled criterion for
  crack nucleation in brittle materials to the {{3D}} case}.
\newblock \bibinfo{journal}{Theoretical and Applied Fracture Mechanics}
  \bibinfo{volume}{74}, \bibinfo{pages}{7--17}.
\newblock \DOIprefix\doi{10.1016/j.tafmec.2014.05.004}.
\bibitem[{Lubomirsky and
  Bouchbinder(2023)}]{lubomirskyQuenchedDisorderInstability2023}
\bibinfo{author}{Lubomirsky, Y.}, \bibinfo{author}{Bouchbinder, E.},
  \bibinfo{year}{2023}.
\newblock \bibinfo{title}{Quenched disorder and instability control dynamic
  fracture in three dimensions}.
\newblock \DOIprefix\doi{10.48550/arXiv.2311.11692},
  \href{http://arxiv.org/abs/2311.11692}{{\tt arXiv:2311.11692}}.
\bibitem[{Mirkhalaf et~al.(2014)Mirkhalaf, Dastjerdi and
  Barthelat}]{mirkhalafOvercomingBrittlenessGlass2014}
\bibinfo{author}{Mirkhalaf, M.}, \bibinfo{author}{Dastjerdi, A.K.},
  \bibinfo{author}{Barthelat, F.}, \bibinfo{year}{2014}.
\newblock \bibinfo{title}{Overcoming the brittleness of glass through
  bio-inspiration and micro-architecture}.
\newblock \bibinfo{journal}{Nature Communications} \bibinfo{volume}{5}.
\bibitem[{M{\"o}ller and Bitzek(2015)}]{mollerInfluenceCrackFront2015}
\bibinfo{author}{M{\"o}ller, J.J.}, \bibinfo{author}{Bitzek, E.},
  \bibinfo{year}{2015}.
\newblock \bibinfo{title}{On the influence of crack front curvature on the
  fracture behavior of nanoscale cracks}.
\newblock \bibinfo{journal}{Engineering Fracture Mechanics}
  \bibinfo{volume}{150}, \bibinfo{pages}{197--208}.
\newblock \DOIprefix\doi{10.1016/j.engfracmech.2015.03.028}.
\bibitem[{Morrissey and Rice(1998)}]{morrisseyCrackFrontWaves1998}
\bibinfo{author}{Morrissey, J.W.}, \bibinfo{author}{Rice, J.R.},
  \bibinfo{year}{1998}.
\newblock \bibinfo{title}{Crack front waves}.
\newblock \bibinfo{journal}{Journal of the Mechanics and Physics of Solids}
  \bibinfo{volume}{46}, \bibinfo{pages}{467--487}.
\bibitem[{Morrissey and Rice(2000)}]{Morrissey.00}
\bibinfo{author}{Morrissey, J.W.}, \bibinfo{author}{Rice, J.R.},
  \bibinfo{year}{2000}.
\newblock \bibinfo{title}{Perturbative simulations of crack front waves}.
\newblock \bibinfo{journal}{Journal of the Mechanics and Physics of Solids}
  \bibinfo{volume}{48}, \bibinfo{pages}{1229--1251}.
\newblock \DOIprefix\doi{10.1016/S0022-5096(99)00069-1}.
\bibitem[{Movchan et~al.(1998)Movchan, Gao and
  Willis}]{movchanPerturbationsPlaneCracks1998}
\bibinfo{author}{Movchan, A.B.}, \bibinfo{author}{Gao, H.},
  \bibinfo{author}{Willis, J.R.}, \bibinfo{year}{1998}.
\newblock \bibinfo{title}{On perturbations of plane cracks}.
\newblock \bibinfo{journal}{International Journal of Solids and Structures}
  \bibinfo{volume}{35}, \bibinfo{pages}{3419--3453}.
\bibitem[{Norris and
  Abrahams(2007)}]{norrisMultiplescalesApproachCrackfront2007}
\bibinfo{author}{Norris, A.N.}, \bibinfo{author}{Abrahams, I.D.},
  \bibinfo{year}{2007}.
\newblock \bibinfo{title}{A multiple-scales approach to crack-front waves}.
\newblock \bibinfo{journal}{Journal of Engineering Mathematics}
  \bibinfo{volume}{59}, \bibinfo{pages}{399--417}.
\bibitem[{Perrin and Rice(1994)}]{perrinDisorderingDynamicPlanar1994}
\bibinfo{author}{Perrin, G.}, \bibinfo{author}{Rice, J.R.},
  \bibinfo{year}{1994}.
\newblock \bibinfo{title}{Disordering of a dynamic planar crack front in a
  model elastic medium of randomly variable toughness}.
\newblock \bibinfo{journal}{Journal of the Mechanics and Physics of Solids}
  \bibinfo{volume}{42}, \bibinfo{pages}{1047--1064}.
\bibitem[{Pons and Karma(2010)}]{ponsHelicalCrackfrontInstability2010}
\bibinfo{author}{Pons, A.J.}, \bibinfo{author}{Karma, A.},
  \bibinfo{year}{2010}.
\newblock \bibinfo{title}{Helical crack-front instability in mixed-mode
  fracture}.
\newblock \bibinfo{journal}{Nature} \bibinfo{volume}{464},
  \bibinfo{pages}{85--89}.
\bibitem[{Ponson(2009)}]{ponsonDepinningTransitionFailure2009}
\bibinfo{author}{Ponson, L.}, \bibinfo{year}{2009}.
\newblock \bibinfo{title}{Depinning {{Transition}} in the {{Failure}} of
  {{Inhomogeneous Brittle Materials}}}.
\newblock \bibinfo{journal}{Physical Review Letters} \bibinfo{volume}{103},
  \bibinfo{pages}{055501}.
\newblock \DOIprefix\doi{10.1103/PhysRevLett.103.055501}.
\bibitem[{Ramanathan(1997)}]{RamanathanThesis}
\bibinfo{author}{Ramanathan, S.}, \bibinfo{year}{1997}.
\newblock \bibinfo{title}{Crack Propagation through Heterogeneous Media}.
\newblock Ph.D. thesis. Harvard University.
\bibitem[{Ramanathan and Fisher(1997)}]{Ramanathan.97}
\bibinfo{author}{Ramanathan, S.}, \bibinfo{author}{Fisher, D.S.},
  \bibinfo{year}{1997}.
\newblock \bibinfo{title}{Dynamics and instabilities of planar tensile cracks
  in heterogeneous media}.
\newblock \bibinfo{journal}{Physical Review Letters} \bibinfo{volume}{79},
  \bibinfo{pages}{877--880}.
\newblock \DOIprefix\doi{10.1103/PhysRevLett.79.877}.
\bibitem[{Ray and Viesca(2017)}]{rayEarthquakeNucleationFaults2017}
\bibinfo{author}{Ray, S.}, \bibinfo{author}{Viesca, R.C.},
  \bibinfo{year}{2017}.
\newblock \bibinfo{title}{Earthquake {{Nucleation}} on {{Faults With
  Heterogeneous Frictional Properties}}, {{Normal Stress}}}.
\newblock \bibinfo{journal}{Journal of Geophysical Research: Solid Earth}
  \bibinfo{volume}{122}, \bibinfo{pages}{8214--8240}.
\newblock \DOIprefix\doi{10.1002/2017JB014521}.
\bibitem[{Rice et~al.(1994)Rice, {Ben-Zion} and
  Kim}]{riceThreedimensionalPerturbationSolution1994}
\bibinfo{author}{Rice, J.R.}, \bibinfo{author}{{Ben-Zion}, Y.},
  \bibinfo{author}{Kim, K.S.}, \bibinfo{year}{1994}.
\newblock \bibinfo{title}{Three-dimensional perturbation solution for a dynamic
  planar crack moving unsteadily in a model elastic solid}.
\newblock \bibinfo{journal}{Journal of the Mechanics and Physics of Solids}
  \bibinfo{volume}{42}, \bibinfo{pages}{813--843}.
\bibitem[{Roch et~al.(2022)Roch, Barras, Geubelle and
  Molinari}]{rochCRackletSpectralBoundary2022}
\bibinfo{author}{Roch, T.}, \bibinfo{author}{Barras, F.},
  \bibinfo{author}{Geubelle, P.H.}, \bibinfo{author}{Molinari, J.F.},
  \bibinfo{year}{2022}.
\newblock \bibinfo{title}{{{cRacklet}}: A spectral boundary integral method
  library for interfacial rupture simulation}.
\newblock \bibinfo{journal}{Journal of Open Source Software}
  \bibinfo{volume}{7}, \bibinfo{pages}{3724}.
\newblock \DOIprefix\doi{10.21105/joss.03724}.
\bibitem[{Roch et~al.(2023)Roch, Lebihain and
  Molinari}]{rochDynamicCrackFrontDeformations2023}
\bibinfo{author}{Roch, T.}, \bibinfo{author}{Lebihain, M.},
  \bibinfo{author}{Molinari, J.F.}, \bibinfo{year}{2023}.
\newblock \bibinfo{title}{Dynamic {{Crack-Front Deformations}} in {{Cohesive
  Materials}}}.
\newblock \bibinfo{journal}{Physical Review Letters} \bibinfo{volume}{131},
  \bibinfo{pages}{096101}.
\newblock \DOIprefix\doi{10.1103/PhysRevLett.131.096101}.
\bibitem[{Sagy et~al.(2004)Sagy, Fineberg and Reches}]{Sagy.04}
\bibinfo{author}{Sagy, A.}, \bibinfo{author}{Fineberg, J.},
  \bibinfo{author}{Reches, Z.}, \bibinfo{year}{2004}.
\newblock \bibinfo{title}{Shatter cones: {{Branched}}, rapid fractures formed
  by shock impact}.
\newblock \bibinfo{journal}{Journal of Geophysical Research: Solid Earth}
  \bibinfo{volume}{109}.
\bibitem[{Schmittbuhl and
  M{\aa}l{\o}y(1997)}]{schmittbuhlDirectObservationSelfAffine1997}
\bibinfo{author}{Schmittbuhl, J.}, \bibinfo{author}{M{\aa}l{\o}y, K.J.},
  \bibinfo{year}{1997}.
\newblock \bibinfo{title}{Direct {{Observation}} of a {{Self-Affine Crack
  Propagation}}}.
\newblock \bibinfo{journal}{Physical Review Letters} \bibinfo{volume}{78},
  \bibinfo{pages}{3888--3891}.
\newblock \DOIprefix\doi{10.1103/PhysRevLett.78.3888}.
\bibitem[{Shaikeea et~al.(2022)Shaikeea, Cui, O'Masta, Zheng and
  Deshpande}]{shaikeeaToughnessMechanicalMetamaterials2022}
\bibinfo{author}{Shaikeea, A.J.D.}, \bibinfo{author}{Cui, H.},
  \bibinfo{author}{O'Masta, M.}, \bibinfo{author}{Zheng, X.R.},
  \bibinfo{author}{Deshpande, V.S.}, \bibinfo{year}{2022}.
\newblock \bibinfo{title}{The toughness of mechanical metamaterials}.
\newblock \bibinfo{journal}{Nature Materials} \bibinfo{volume}{21},
  \bibinfo{pages}{297--304}.
\newblock \DOIprefix\doi{10.1038/s41563-021-01182-1}.
\bibitem[{Sharon et~al.(2001)Sharon, Cohen and
  Fineberg}]{sharonPropagatingSolitaryWaves2001}
\bibinfo{author}{Sharon, E.}, \bibinfo{author}{Cohen, G.},
  \bibinfo{author}{Fineberg, J.}, \bibinfo{year}{2001}.
\newblock \bibinfo{title}{Propagating solitary waves along a rapidly moving
  crack front}.
\newblock \bibinfo{journal}{Nature} \bibinfo{volume}{410},
  \bibinfo{pages}{68--71}.
\bibitem[{Sharon et~al.(2002)Sharon, Cohen and Fineberg}]{sharon2002crack}
\bibinfo{author}{Sharon, E.}, \bibinfo{author}{Cohen, G.},
  \bibinfo{author}{Fineberg, J.}, \bibinfo{year}{2002}.
\newblock \bibinfo{title}{Crack front waves and the dynamics of a rapidly
  moving crack}.
\newblock \bibinfo{journal}{Physical Review Letters} \bibinfo{volume}{88},
  \bibinfo{pages}{85503}.
\bibitem[{Sharon and Fineberg(1999)}]{Sharon.99}
\bibinfo{author}{Sharon, E.}, \bibinfo{author}{Fineberg, J.},
  \bibinfo{year}{1999}.
\newblock \bibinfo{title}{Confirming the continuum theory of dynamic brittle
  fracture for fast cracks}.
\newblock \bibinfo{journal}{Nature} \bibinfo{volume}{397},
  \bibinfo{pages}{333--335}.
\newblock \DOIprefix\doi{10.1038/16891}.
\bibitem[{Stanchits et~al.(2015)Stanchits, Burghardt and
  Surdi}]{stanchitsHydraulicFracturingHeterogeneous2015}
\bibinfo{author}{Stanchits, S.}, \bibinfo{author}{Burghardt, J.},
  \bibinfo{author}{Surdi, A.}, \bibinfo{year}{2015}.
\newblock \bibinfo{title}{Hydraulic {{Fracturing}} of {{Heterogeneous Rock
  Monitored}} by {{Acoustic Emission}}}.
\newblock \bibinfo{journal}{Rock Mechanics and Rock Engineering}
  \bibinfo{volume}{48}, \bibinfo{pages}{2513--2527}.
\newblock \DOIprefix\doi{10.1007/s00603-015-0848-1}.
\bibitem[{Steinhardt and
  Rubinstein(2022)}]{steinhardtHowMaterialHeterogeneity2022}
\bibinfo{author}{Steinhardt, W.}, \bibinfo{author}{Rubinstein, S.M.},
  \bibinfo{year}{2022}.
\newblock \bibinfo{title}{How {{Material Heterogeneity Creates Rough
  Fractures}}}.
\newblock \bibinfo{journal}{Physical Review Letters} \bibinfo{volume}{129},
  \bibinfo{pages}{128001}.
\newblock \DOIprefix\doi{10.1103/PhysRevLett.129.128001}.
\bibitem[{Svetlizky and Fineberg(2014)}]{svetlizkyClassicalShearCracks2014}
\bibinfo{author}{Svetlizky, I.}, \bibinfo{author}{Fineberg, J.},
  \bibinfo{year}{2014}.
\newblock \bibinfo{title}{Classical shear cracks drive the onset of dry
  frictional motion}.
\newblock \bibinfo{journal}{Nature} \bibinfo{volume}{509},
  \bibinfo{pages}{205--208}.
\bibitem[{Vasoya et~al.(2016)Vasoya, Unni, Leblond, Lazarus and
  Ponson}]{vasoyaFiniteSizeGeometrical2016}
\bibinfo{author}{Vasoya, M.}, \bibinfo{author}{Unni, A.B.},
  \bibinfo{author}{Leblond, J.B.}, \bibinfo{author}{Lazarus, V.},
  \bibinfo{author}{Ponson, L.}, \bibinfo{year}{2016}.
\newblock \bibinfo{title}{Finite size and geometrical non-linear effects during
  crack pinning by heterogeneities: {{An}} analytical and experimental study}.
\newblock \bibinfo{journal}{Journal of the Mechanics and Physics of Solids}
  \bibinfo{volume}{89}, \bibinfo{pages}{211--230}.
\bibitem[{Willis(2013)}]{willisCrackFrontPerturbations2013}
\bibinfo{author}{Willis, J.R.}, \bibinfo{year}{2013}.
\newblock \bibinfo{title}{Crack front perturbations revisited}.
\newblock \bibinfo{journal}{International Journal of Fracture}
  \bibinfo{volume}{184}, \bibinfo{pages}{17--24}.
\bibitem[{Willis and Movchan(1995)}]{willisDynamicWeightFunctions1995}
\bibinfo{author}{Willis, J.R.}, \bibinfo{author}{Movchan, A.B.},
  \bibinfo{year}{1995}.
\newblock \bibinfo{title}{Dynamic weight functions for a moving crack. {{I}}.
  {{Mode I}} loading}.
\newblock \bibinfo{journal}{Journal of the Mechanics and Physics of Solids}
  \bibinfo{volume}{43}, \bibinfo{pages}{319--341}.
\bibitem[{Willis and
  Movchan(1997)}]{willisThreedimensionalDynamicPerturbation1997}
\bibinfo{author}{Willis, J.R.}, \bibinfo{author}{Movchan, A.B.},
  \bibinfo{year}{1997}.
\newblock \bibinfo{title}{Three-dimensional dynamic perturbation of a
  propagating crack}.
\newblock \bibinfo{journal}{Journal of the Mechanics and Physics of Solids}
  \bibinfo{volume}{45}, \bibinfo{pages}{591--610}.
\bibitem[{Xia et~al.(2012)Xia, Ponson, Ravichandran and
  Bhattacharya}]{xiaTougheningAsymmetryPeeling2012}
\bibinfo{author}{Xia, S.}, \bibinfo{author}{Ponson, L.},
  \bibinfo{author}{Ravichandran, G.}, \bibinfo{author}{Bhattacharya, K.},
  \bibinfo{year}{2012}.
\newblock \bibinfo{title}{Toughening and asymmetry in peeling of heterogeneous
  adhesives}.
\newblock \bibinfo{journal}{Physical Review Letters} \bibinfo{volume}{108},
  \bibinfo{pages}{1--5}.
\newblock \DOIprefix\doi{10.1103/PhysRevLett.108.196101},
  \href{http://arxiv.org/abs/1203.3634}{{\tt arXiv:1203.3634}}.

\end{thebibliography}
\end{document}